# KATEGORISASI DOKUMEN WEB SECARA OTOMATIS BERDASARKAN FOLKSONOMY MENGGUNAKAN MULTINOMIAL NAIVE BAYES CLASSIFIER

# ( AUTOMATIC FOLKSONOMY CATEGORIZATION OF WEB DOCUMENTS USING MULTINOMIAL NAIVE BAYES CLASSIFIER )

**TUGAS AKHIR**

Diajukan Sebagai Syarat Kelulusan

Pendidikan Program Sarjana

Jurusan Teknik Informatika

Sekolah Tinggi Teknologi TELKOM

Oleh

**Hendy Irawan**

113010043

STTTELKOM

# Abstrak

*Folksonomy* merupakan metode kategorisasi dokumen yang tidak hierarkis, menyamaratakan kedudukan setiap kategori, dan judul kategori ditentukan secara bebas oleh siapa saja yang memasukkan sebuah dokumen di dalam kategori-kategori tersebut. Pembuatan kategorisasi dilakukan secara otomatis pada saat dokumen dimasukkan, yaitu dengan cara mengetikkan daftar kategori yang kira-kira cocok untuk dokumen tersebut. Situs del.icio.us (http://del.icio.us) merupakan salah satu *social bookmarking site* terpopuler yang menggunakan *folksonomy*.

Penggunaan *folksonomy*, meski sangat mudah, juga mempunyai beberapa kelemahan, yaitu penggunaan *tag* yang berbeda-beda untuk konsep yang sama, penggunaan *tag* yang sama untuk konsep yang berbeda-beda, tidak adanya pengendalian mutu, dan lain-lain. Di sini penulis mencoba memberikan solusi untuk sebagian masalah tersebut yaitu dengan cara menganalisa isi dari dokumen Web yang ditunjuk dan mengkategorisasikan link tersebut secara otomatis ke beberapa *tag* menggunakan multinomial naive Bayes classifier.

Bayes classifier bekerja berdasarkan sekumpulan bukti (*evidence*) dan kelas (*class*). Dengan melakukan pelatihan (*training*) terhadap sebagian data sampel, dapat ditentukan probabilitas kepastian (*likelihood probability*) dari sebuah bukti jika diberikan kelas tertentu. Bayes classifier juga menggunakan probabilitas sebelumnya (*prior probability*) dari sebuah kelas, yang perhitungannya dapat didasarkan dari sampel data tersebut. Dari analisa sampel data tersebut, jika diberikan sebuah dokumen baru yang terdiri dari sekumpulan bukti, probabilitas setiap kelas terhadap dokumen tersebut (*posterior probability*) dapat ditentukan.

Sistem ini diimplementasikan menggunakan PHP 5, Apache, dan MySQL. Kesimpulan yang didapatkan dari penelitian ini adalah metode Bayes dapat digunakan untuk melakukan kategorisasi dokumen secara otomatis maupun sebagai alat bantu untuk kategorisasi manual.

# Abstract

Folksonomy is a non-hierarchical document categorizing system, that treats every category in a flat manner, dan every category is entered freely by anyone who submitted a document in these categories. Categorization is done automatically at the time a document is submitted, by entering the list of categories that best fit the document. del.icio.us (http://del.icio.us) site is one of the most popular social bookmarking sites that uses folksonomy.

Usage of folksonomy, although very easy, also has its weaknesses, such as use of different tags for the same concept, use of the same tag for different concepts, no quality control, etc. We try to provide a solution for some of these problems by analyzing Web documents' contents and categorizing them automatically using multinomial naive Bayes algorithm.

Bayes classifier works by using a set of evidences and a set of classes. By training the system using sample data, we can determine the probability of an evidence given a particular class. Bayes classifier also uses prior probability of a class, which can be calculated from sample data. From these analysis, when given a new document which is formed by a set of evidences (words), the probabilities of each class given that document (posterior probabilities) can be determined.

This system is implemented using PHP 5, Apache, and MySQL. The conclusion from building this system is that the Bayes method can be used to automatically categorize documents and also as an assistive tool for manual categorization.

# Daftar Isi

# Bab I
# Pendahuluan

## 1.1 Latar Belakang

Perkembangan World Wide Web dalam 10 tahun terakhir ini semakin pesat, dan sekarang keberadaan Internet sudah tidak bisa diabaikan lagi. Dengan banyaknya informasi yang tersedia di Internet, berbagai layanan telah dikembangkan untuk membantu pengguna Internet dalam menemukan informasi yang mereka inginkan.

Layanan yang paling populer adalah mesin pencari atau *search engine*, misalnya Google (http://www.google.com), Ask Jeeves (http://www.ask.com), dan MSN Search (http://www.msnsearch.com), yang dapat digunakan untuk mencari informasi berdasarkan kata-kata kunci tertentu. Meski sangat berguna, kadang-kadang kita ingin mencari informasi berdasarkan kategori tertentu. Untuk itu ada layanan direktori, misalnya Yahoo! (http://www.yahoo.com) dan dmoz.org Open Directory (http://www.dmoz.org), yang berisi daftar situs Web berdasarkan klasifikasi tertentu. Layanan direktori ini sebenarnya sangat berguna, namun memiliki banyak kelemahan, di antaranya adalah kurang terupdate, struktur kategori yang statis, dan kurangnya staf untuk merawat ribuan kategori dan subkategori yang terdapat dalam sebuah layanan direktori.

Kategori situs yang akhir-akhir ini populer adalah *social networking sites*, misalnya Friendster (http://www.friendster.com) dan GaulDong (http://www.gauldong.net). Pesatnya perkembangan *social sites* membuat banyak layanan lain bermunculan, salah satunya adalah del.icio.us (http://del.icio.us) yang merupakan *social bookmarking site* berdasarkan *folksonomy*. Setiap pengguna del.icio.us bebas mengirim link/bookmark ke sebuah situs yang dikategorikan ke satu atau lebih *tag* (istilah del.icio.us untuk kategori). *Folksonomy* merupakan teknik klasifikasi yang tidak hierarkis, melainkan semua kategori disamaratakan, dan kategori dibuat secara bebas berdasarkan masukan dari pengguna.

Penggunaan *folksonomy*, meski sangat mudah, juga mempunyai beberapa kelemahan, yaitu penggunaan *tag* yang berbeda-beda untuk konsep yang sama, penggunaan *tag* yang sama untuk konsep yang berbeda-beda, tidak adanya pengendalian mutu, dan lain-lain. Di sini penulis mencoba memberikan solusi untuk sebagian masalah tersebut yaitu dengan cara menganalisa isi dari dokumen Web yang ditunjuk dan mengkategorisasikan link tersebut secara otomatis ke beberapa *tag* menggunakan multinomial naive Bayes classifier untuk keperluan ini.

## 1.2 Perumusan Masalah

Permasalahan yang dijadikan objek penelitian tugas akhir ini adalah menitikberatkan pada analisa penggunaan multinomial naive Bayes classifier untuk kategorisasi dokumen Web secara otomatis. Dari penelitian ini diharapkan dapat diketahui bagaimana performansi algoritma tersebut dalam melakukan kategorisasi otomatis dokumen Web.

## 1.3 Tujuan

Tujuan atau hasil akhir yang ingin dicapai dari tugas akhir ini adalah:

1. Mengimplementasikan sebuah sistem kategorisasi dokumen Web dengan fasilitas minimal, bernama Gado-gado.
2. Menerapkan multinomial naive Bayes classifier untuk kategorisasi dokumen secara otomatis.
3. Melakukan analisa terhadap akurasi kategorisasi otomatis yang dilakukan dibandingkan dengan kategorisasi secara manual, serta analisa performansi dan efisiensi.

## 1.4 Batasan Masalah

Untuk menghindari meluasnya materi pembahasan tugas akhir ini, maka penulis membatasi permasalahan dalam tugas akhir ini hanya mencakup hal-hal berikut:

1. Aplikasi akan dibangun dengan menggunakan PHP 5.0.3 sebagai web scripting language, Apache 2.0 sebagai web server, MySQL sebagai database management system, dan Windows XP Professional sebagai operating system.
2. Aktivitas yang dilakukan dalam sistem ini meliputi login user, penambahan link, menampilkan daftar link dalam sebuah tag, dan melakukan uji coba.
3. Dokumen yang dapat diterima hanya dalam bahasa Inggris (variasi apa pun) atau Indonesia (tidak menggunakan dua bahasa dalam satu dokumen), dengan encoding berikut: ISO-8859-1, WIN-1252, atau UTF-8.
4. Ukuran keseluruhan database maksimal 50 MB.

## 1.5 Metodologi Penyelesaian Masalah

Berikut ini adalah metodologi penyelesaian masalah yang dipergunakan dalam tugas akhir ini:

1. Studi literatur

   Bertujuan untuk mempelajari dasar teori dari literatur-literatur tentang :

- Folksonomy
- Indexing
- Inverted index
- Klasifikasi teks
- Teori probabilitas
- Bayes theorem
- Naive Bayes classifier
- Multinomial naive Bayes classifier

2. Pengumpulan data untuk inputan
3. Studi perancangan perangkat lunak

    Bertujuan untuk menentukan metodologi pengembangan perangkat lunak yang digunakan dengan menggunakan metode terstruktur dan melakukan perancangan sistem.

4. Pembuatan perangkat lunak

    Bertujuan untuk melakukan implementasi metode pada perangkat lunak sesuai dengan analisa perancangan yang telah dilakukan.

5. Pengujian perangkat lunak
6. Analisa terhadap hasil pengujian
7. Pengambilan kesimpulan dan penyusunan laporan

## *1.6 Sistematika Penulisan*

**BAB I PENDAHULUAN (INTRODUCTION)**

Berisi latar belakang, perumusan masalah, tujuan pembahasan., metodologi penyelesaian masalah dan sistematika penulisan.

**BAB II LANDASAN TEORI (FUNDAMENTAL THEORIES)**

Penjelasan mengenai *folksonomy*, *indexing*, *inverted index*, *text classifier*, teori probabilitas, Bayes *theorem*, *naive* Bayes, *multinomial naive* Bayes.

**BAB III ANALISIS DAN PERANCANGAN SISTEM (SYSTEM ANALYSIS AND DESIGN)**

Membahas tentang perancangan awal sistem dengan metode Test-Driven Development (TDD) menggunakan bahasa pemodelan UML.

**BAB IV IMPLEMENTASI DAN PENGUJIAN (IMPLEMENTATION AND TESTING)**

Menyajikan hasil pengujian dan analisa terhadap kategorisasi dokumen Web secara otomatis berdasarkan folksonomy menggunakan multinomial naive Bayes classifier.

**BAB V KESIMPULAN DAN SARAN (CONCLUSIONS AND SUGGESTIONS)**

Berisi kesimpulan dan saran pengembangan.

# Bab II
# Landasan Teori

## *2.1 Sistem Klasifikasi*

Pada subbab ini akan diberikan deskripsi dan perbandingan mengenai beberapa sistem klasifikasi yang umum dipakai, yaitu sistem hierarkis (*hierarchical classification systems*) dan *faceted classification systems*, serta sistem klasifikasi yang dipakai untuk pengerjaan Tugas Akhir ini yaitu *folksonomy*.

### 2.1.1 Hierarchical Systems

Sistem hierarkis atau pohon (*tree*) merupakan sistem klasifikasi yang paling tua dan paling umum dipakai. Sistem klasifikasi ini terdiri dari *node-node* dan dimulai dari sebuah akar (*root*) yang mempunyai beberapa cabang (*branch*). Setiap cabang atau *node* dapat mempunyai cabang-cabang lain. *Node* yang tidak memiliki cabang dinamakan *leaf node*. Tergantung dari penggunaan klasifikasinya, dokumen atau *item* dapat dimasukkan dalam *node* apa pun atau hanya *leaf node* saja.

Untuk menavigasi sistem hierarkis, seseorang mulai dari *root node* dan memilih salah satu cabang yang diinginkan atau yang kira-kira paling sesuai dengan item yang dicari. Penjelajahan diteruskan sampai menemukan *leaf node* atau dokumen/*item* yang diinginkan.

Sistem hierarkis merupakan sistem klasifikasi yang paling banyak digunakan, karena jelas dan mudah pemakaiannya serta intuitif. Namun bukan tanpa kelemahan, karena sistem ini menerapkan klasifikasi yang baku. Misalnya, taksonomi untuk hewan dibagi menjadi dua bagian besar yaitu invertebrata (tidak bertulang belakang) dan vertebrata (bertulang belakang). Seseorang yang tidak mempunyai pengetahuan cukup di bidang ini akan kesulitan untuk mencari hewan yang diinginkan. Kesulitan lainnya adalah klasifikasi yang tidak intuitif seperti paus yang bukan diklasifikasikan sebagai ikan tetapi mamalia. Ada juga klasifikasi yang sulit misalnya jenis *velvetworm*, yang mempunyai karakteristik seperti *arthropoda* tetapi juga memiliki karakteristik *annelida*. Kadangkala multikategori dapat digunakan untuk mengatasi kesulitan-kesulitan tersebut.

Berbagai sistem yang menggunakan sistem klasifikasi hierarkis di antaranya:

- Klasifikasi/taksonomi makhluk hidup (famili, genus, spesies, dsb.)
- Dewey Decimal System untuk klasifikasi buku di perpustakaan
- Struktur direktori di *file system* PC

Dalam kaitannya dengan Tugas Akhir ini, situs-situs web yang menggunakan sistem klasifikasi hierarkis untuk mengklasifikasikan situs web antara lain:

- Yahoo (http://www.yahoo.com)
- Open Directory (http://www.dmoz.org)

### 2.1.2 Faceted Systems

*Faceted classification systems* mempunyai cara navigasi yang hampir sama dengan sistem hierarkis, namun implementasinya berbeda. Sistem ini ditemukan pada tahun 1930-an oleh Shiyali Ranganathan, seorang pustakawan dari India, dengan memberikan sebuah set parameter tertentu untuk objek-objek yang terdapat di dalamnya. Sebagai contoh:

- Untuk objek komputer, maka daftar parameternya adalah harga, kapasitas, kecepatan, jenis prosesor, dan sebagainya
- Untuk objek makanan, maka daftar parameternya adalah asal daerah, harga, basah/kering, makanan pembuka/utama/penutup, dan sebagainya

Dalam menavigasi klasifikasi ini, pilihan yang diberikan mirip dengan klasifikasi hierarkis, namun orang dapat memilih parameter apa pun yang diinginkan, sambil terus memperinci kriterianya. Sebagai contoh, dengan menggunakan taksonomi hewan yang baku, kita hanya dihadapkan pada dua pilihan pada awal navigasi, yaitu invertebrata atau vertebrata. Dengan klasifikasi *faceted*, kita bisa memilih berdasarkan misalnya jumlah kaki, organ pernafasan, atau kriteria yang lain.

Situs web yang menggunakan klasifikasi *faceted* antara lain:

- Amazon.com (http://www.amazon.com)
- Computers4U.com (http://www.computers4u.com)

Kebanyakan situs yang menggunakan klasifikasi *faceted* adalah situs e-commerce, ini bisa dimaklumi karena mereka ingin memudahkan pengguna dalam mencari produk yang akan dibeli.

### 2.1.3 Folksonomy

*Folksonomy* adalah pendatang baru di sistem klasifikasi, bahkan namanya pun secara tidak formal diambil dari kebiasaan orang ("folks") untuk mengkategorikan sebuah *item* dalam taksonomi tertentu. Konsep *folksonomy* sebenarnya sudah ada sejak lama, yaitu dengan mengasosiasikan sebuah item dengan beberapa *tag* (disebut juga dengan *keyword*) tertentu, namun popularitasnya baru mencuat akhir-akhir ini dengan adanya situs del.icio.us

(http://del.icio.us, sebuah situs *social bookmarking*) yang mendayagunakan *folksonomy* ini secara efektif sehingga berguna bagi para pengguna situs ini. del.icio.us tidak menerapkan sistem yang baku melainkan bebas, jadi pengguna dapat mengetikkan kata apa pun sebagai *tag* dan sistem akan menerimanya.

Sistem *folksonomy* yang praktis dan mudah dipakai ini terbukti sukses dan disukai oleh banyak orang karena tidak merepotkan namun cukup berguna, terutama di situs-situs komunitas seperti del.icio.us, Flickr (http://www.flickr.com, situs *social photoblogging*), Technorati (http://www.technorati.com), dan lain-lain. Tentu saja sistem *folksonomy* yang sangat sederhana ini mempunyai banyak sekali kelemahan, di antaranya:

- Satu *tag* dapat mengandung beberapa arti, tergantung konteks. Misalnya, tag *apple* bisa digunakan untuk menyebut buah apel tapi bisa juga digunakan untuk perusahaan Apple Computing.
- Satu arti yang sama bisa mempunyai beberapa *tag* yang berbeda. Contohnya *mac* dan *macintosh* mempunyai arti yang sama, tapi sistem tetap membedakan kedua tag tersebut.
- Satu kata memiliki ejaan yang berbeda-beda. Contohnya jogja, yogya, dan yogyakarta.
- Batasan lingkup sebuah tag tidak jelas. Apabila sebuah bookmark termasuk dalam *tag PHP*, bisa dipastikan dokumen tersebut termasuk dalam *tag programming*, apalagi *computing*. Namun sangat tidak praktis untuk mencantumkan semua tag yang cocok dengan bookmark tersebut karena jumlahnya sangat banyak.

Sistem yang dikembangkan dalam Tugas Akhir ini, yang kami beri nama Gado-gado, merupakan situs tiruan del.icio.us dan menggunakan klasifikasi *folksonomy*, tapi menerapkan algoritma Bayes untuk membantu mengatasi kelemahan-kelemahan *folksonomy*.

## *2.2 Bayes Classifier*

Pada subbab ini akan dijelaskan mengenai Bayes classifier mulai dari teori dasar statistik yang berkaitan varian yang kami gunakan yaitu multinomial naive Bayes.

### 2.2.1 Teori Probabilitas dan Statistika

Teorema Bayes adalah teorema statistik oleh karena itu semua hukum dasar statistik berlaku di sini. Teorema Bayes juga merupakan teorema probabilitas sehingga teori-teori probabilitas berlaku juga.

Beberapa teori dan notasi dasar yang dipakai di sini antara lain:

| | |
|---|---|
| $P(A)$ | Berarti probabilitas terjadinya A. |
| $P(A \mid B)$ | Adalah probabilitas terjadinya A apabila diketahui B. |
| $P(A \& B)$ | Adalah probabilitas terjadinya A dan B secara bersamaan. |

Di mana:

$$P(A \mid B) = \frac{P(A \& B)}{P(B)}$$

Beberapa contoh notasi lain:

| | |
|---|---|
| $\lvert d \rvert$ | Jumlah elemen di dalam himpunan d. |
| $\prod_{t=1}^{\lvert V \rvert} p(W = w_t \mid c_j)$ | Untuk setiap elemen di dalam V dari t=1 sampai jumlah elemen V, hitung probabilitas W bila W adalah $w_t$ dan diketahui $c_j$, dan kalikan hasilnya. |

### 2.2.2 Teorema Bayes

Teorema Bayes diturunkan dari teori dasar probabilitas sebagai berikut:

$$P(A \mid B) = \frac{P(A \& B)}{P(B)} \qquad (1)$$

Dengan menggunakan aljabar sederhana kita dapat merumuskannya sebagai berikut:

$$P(A \& B) = P(A \mid B) \times P(B) \qquad (2)$$

Nilai di sebelah kanan dapat juga dihitung menggunakan A sebagai kondisinya:

$$P(A \& B) = P(B \mid A) \times P(A) \qquad (3)$$

Dari persamaan (2) dan persamaan (3) dapat diturunkan persamaan sebagai berikut:

$$P(A \mid B) \times P(B) = P(B \mid A) \times P(A) \qquad (4)$$

Persamaan tersebut apabila disederhanakan akan menghasilkan teorema Bayes yaitu:

$$P(A \mid B) = \frac{P(B \mid A) \times P(A)}{P(B)}$$

Dari persamaan tersebut, kita bisa menghitung P(A|B) apabila diketahui P(B|A), P(A), dan P(B).

### 2.2.3 Bayes Classifier

Teorema Bayes menyatakan bahwa apabila kita mengetahui:

| | |
|---|---|
| $P(d \mid c_j)$ | Probabilitas suatu dokumen termasuk dalam suatu kelas tertentu, di mana j mulai dari 1 sampai jumlah kelas yang ada. |

dan kita juga mengetahui (atau mensintesis):

| | |
|---|---|
| $P(c_j)$ | Probabilitas prior kelas-kelas yang ada (*class prior probabilities*). |

maka kita dapat menghitung dari masukan-masukan di atas:

| | |
|---|---|
| $P(d) = \sum_{j=1}^{\lvert c \rvert} P(d \mid c_j) \times P(c_j)$ | Probabilitas sebuah dokumen. |

dan menghasilkan probabilitas yang kita inginkan yaitu probabilitas sebuah kelas apabila diketahui sebuah dokumen beserta probabilitasnya:

| | |
|---|---|
| $P(c_j \mid d) = \dfrac{P(d \mid c_j) \times P(c_j)}{P(d)}$ | Probabilitas kelas $c_j$ jika diketahui dokumen d. |

Ilustrasinya adalah sebagai berikut:

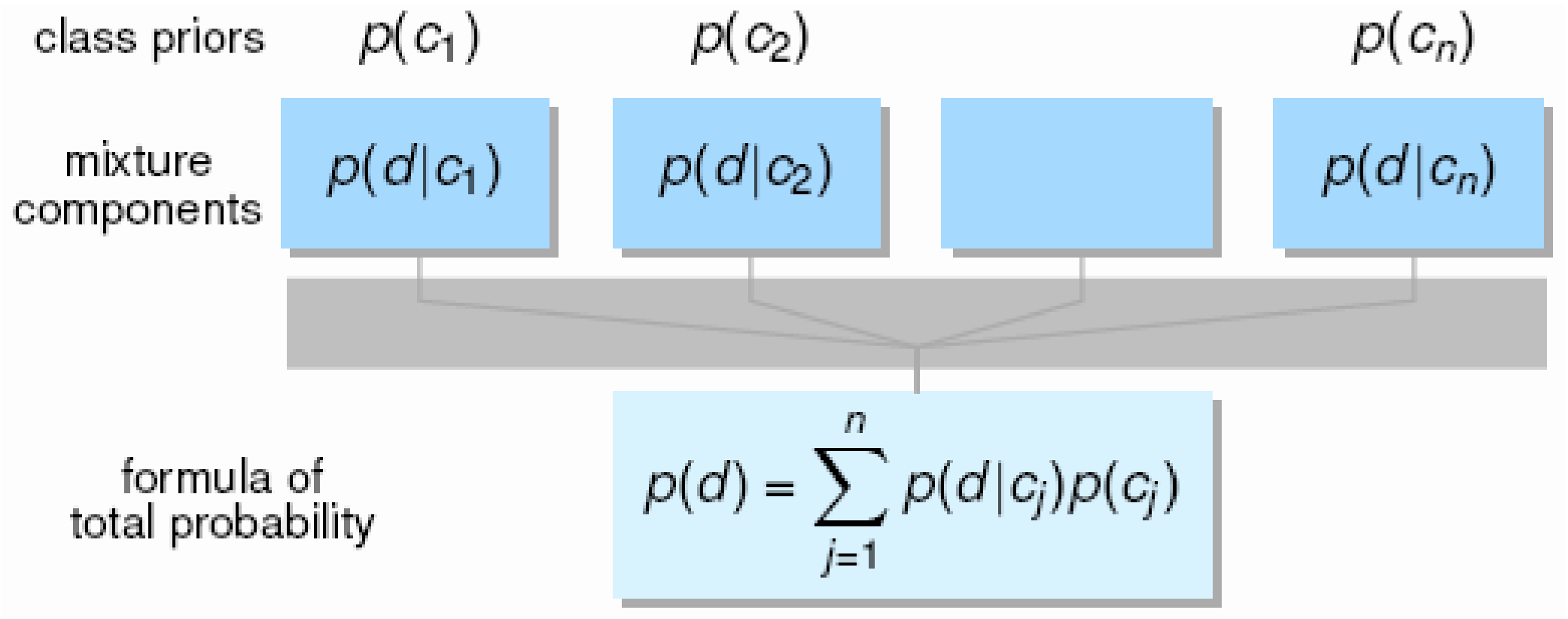

Di mana:

$$\text{Bayes' Rule:} \quad p(c_j \mid d) = \frac{p(d \mid c_j)\, p(c_j)}{p(d)}$$

Untuk menentukan kelas yang paling sesuai untuk dokumen tersebut dengan memilih kelas dengan probabilitas paling tinggi:

$$\text{Classification:} \quad c^*(d) = \operatorname{argmax}_{c_j} p(c_j \mid d)$$

### 2.2.4 Binary Independence Model

*Binary independence model* adalah salah satu model *Bayes classifier* yang umum dipakai dan menggunakan pemetaan satu-satu antara kemunculan suatu kata dalam sebuah dokumen dengan probabilitasnya. Model ini tidak digunakan dalam pengerjaan Tugas Akhir ini dan kami cantumkan hanya sebagai perbandingan teori dan untuk memperjelas teori model multinomial.

Pada binary independence model, sebuah variabel random $W_t$ memodelkan kemunculan kata $w_t$ di dalam dokumen d:

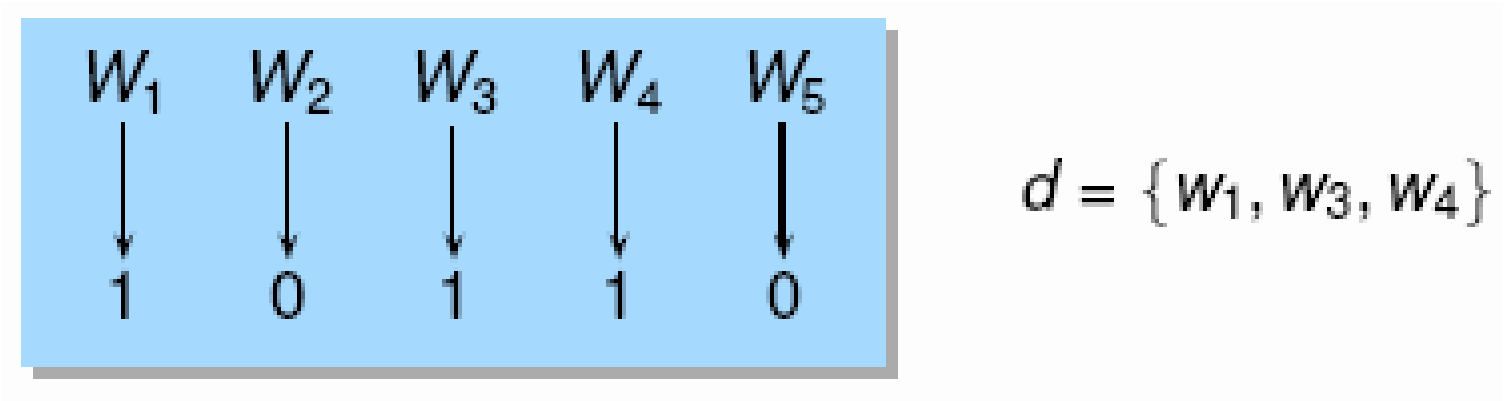

Dari ilustrasi di atas maka perhitungan probabilitas dokumen d jika diketahui $c_j$ adalah sebagai berikut (asumsi setiap probabilitas kata adalah independen):

$$p(d|c_j) = p(W_1 = 1|c_j)p(W_2 = 0|c_j)p(W_3 = 1|c_j)p(W_4 = 1|c_j)p(W_5 = 0|c_j)$$

Binary independence model tidak cocok digunakan untuk klasifikasi teks dikarenakan alasan-alasan sebagai berikut:

- Sebagian besar dokumen adalah *sparse document*, di mana sebagian besar *term* (kata) tidak pernah muncul di dokumen tersebut. Penggunaan binary independence model memberikan pengaruh lebih besar bagi ketiadaan suatu kata di dalam dokumen dibandingkan kata-kata yang muncul dalam dokumen.
- Binary independence model tidak memodelkan frekuensi kemunculan kata dalam dokumen.

Oleh karena itu sistem Gado-gado dalam Tugas Akhir ini menggunakan multinomial model.

### 2.2.5 Multinomial Model

Di dalam model multinomial sebuah variabel random W memodelkan kemunculan kata-kata dalam vocabulary V:

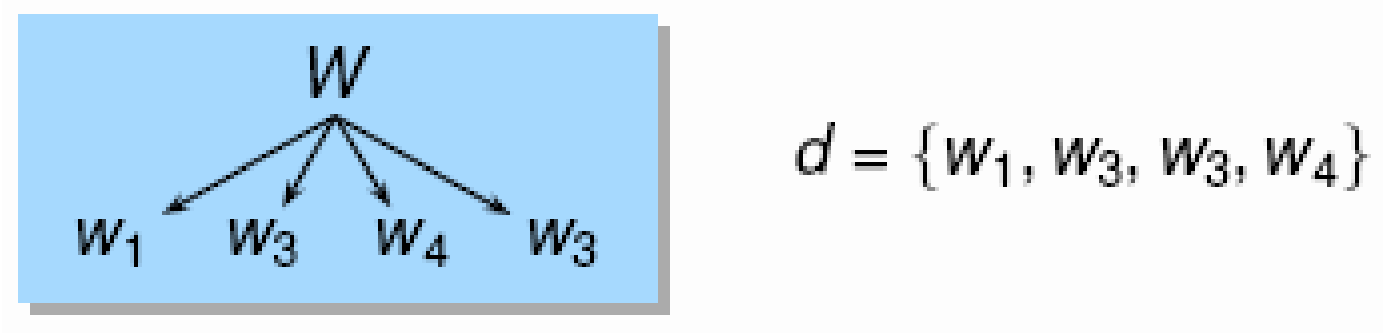

Maka probabilitas dokumen tersebut jika diketahui kelas $c_j$ adalah (asumsi probabilitas setiap kata adalah independen):

$$p(d|c_j) = p(|d| = 4) \cdot 4! \cdot \frac{p(W = w_1|c_j)}{1!} \frac{p(W = w_3|c_j)^2}{2!} \frac{p(W = w_4|c_j)}{1!}$$

Untuk menghitung setiap probabilitas kita dapat menggunakan rumus sebagai berikut:

$$\hat{p}(W = w_t|c_j) = \frac{1 + \#words(c_j, w_t)}{|V| + \#words(c_j)}$$

Pada rumus tersebut dilakukan *smoothing* untuk memberikan nilai probabilitas yang cukup baik apabila jumlah kata sangat kecil (mendekati nol).

### 2.2.6 Multinomial Naive Bayes Classifier

Algoritma klasifikasi teks ini menggabungkan teorema-teorema di atas untuk diterapkan pada permasalahan klasifikasi teks secara otomatis. Algoritma ini disebut “naive” karena mengasumsikan independensi di antara kemunculan kata-kata di dokumen, tanpa memperhitungkan urutan kata dan informasi konteks dalam kalimat atau dokumen secara umum.

### 2.2.7 Perbandingan Sistem ini dengan Text Classifier Lain

Multinomial Naive Bayes banyak dipakai untuk keperluan klasifikasi teks secara umum dan untuk spam filtering. Namun sistem ini mempunyai keunikan dalam penggunaan sebagai berikut:

**Tabel 1. Perbandingan sistem ini dengan text classifier lain**

| *Text Classifier lain (misalnya spam filtering)* | *Sistem ini* |
|---|---|
| Training data yang reliable | Training data yang tidak reliable |
| Satu kategori per dokumen | Beberapa kategori per dokumen |
| Setiap kategori independen (tidak berhubungan) | Banyak kategori yang relevan dan berhubungan |
| Kategori benar-benar disamaratakan | Susunan kategori mempunyai karakteristik hierarkis (ada kategori-kategori yang lebih spesifik dari yang lain) |
| Jumlah kategori sedikit | Jumlah kategori banyak (bahkan sangat banyak) |
| Satu bahasa untuk semua dokumen | Berbagai dokumen dengan berbagai *character encoding* dalam satu database |
| Kategori dalam satu bahasa | Kategori bisa dalam berbagai bahasa |
| Tidak ada kategori yang terduplikasi | Banyak kategori yang relevan bahkan sinonim |
| Jumlah dokumen relatif lebih sedikit | Banyak dokumen |
| Jumlah kata relatif lebih sedikit | Jumlah kata sangat banyak |
| Hanya digunakan untuk keperluan penelitian | Gado-gado rencananya akan dirilis sebagai situs web yang berjalan sepenuhnya |

| *Text Classifier lain (misalnya spam filtering)* | *Sistem ini* |
|---|---|
| Kecepatan tidak terlalu penting (bisa dilakukan analisa secara offline) | Kecepatan sangat mempengaruhi |
| Kapasitas hard disk tidak penting | Kapasitas hard disk penting karena resource di server sangat terbatas |
| Analisa secara offline | Analisa secara kontinu dan online |
| Kategori secara otomatis namun statis | Kategorisasi otomatis yang dinamis, bisa berubah-ubah tergantung training data |
| Kategori tidak bertabrakan | Banyak kategori yang bertabrakan (*overlap* dengan kategori lain) |

Dari gambaran di atas, bisa disimpulkan bahwa pengembangan sistem Gado-gado merupakan tantangan yang jauh lebih berat daripada klasifikasi teks pada umumnya.

## *2.3 Full-text Indexing*

Full-text indexing biasanya digunakan dalam Information Retrieval systems untuk menyimpan dan mencari dokumen berdasarkan *query* yang berupa kata-kata tertentu. Sebelum proses *query* dokumen-dokumen yang ingin diproses harus diindex terlebih dahulu dengan proses yang disebut *full-text indexing*.

Pada sistem Tugas Akhir ini *full-text indexing* digunakan untuk memproses dan melatih sistem menggunakan algoritma multinomial naive bayes. Meski penggunaan utamanya bukan untuk searching tetapi konsep dan teori indexing tetap dipakai dan berlaku di sini.

### 2.3.1 Indexing

Indexing adalah suatu metode untuk mempercepat pencarian data dengan cara membagi domain data menjadi beberapa bagian atau blok dan membuat suatu tabel (*index*) yang berisi pemetaan antara data yang diinginkan dengan lokasi blok data tersebut. Hal ini akan sangat meningkatkan performansi pencarian data karena sistem tidak harus menelusuri data dari awal tapi hanya membaca index-nya saja. Di lain sisi hal ini juga memperumit dan memperlambat proses update data karena setiap kali data ditambah, diubah, atau dihapus, index juga harus di-update.

### 2.3.2 Inverted Index

Inverted index adalah jenis index yang digunakan pada full-text indexing. Disebut *inverted* karena index ini menyimpan pemetaan antara sebuah kata dengan daftar dokumen yang berisi kata tersebut. Contoh inverted index adalah sebagai berikut:

**Tabel 2. Contoh inverted index**

| *Kata* | *Daftar dokumen* |
|---|---|
| elephant | 2, 5, 7, 9 |
| horse | 1, 9, 11 |
| dolphin | 4, 5, 8 |

Dalam contoh tabel di atas, kata *elephant* terdapat dalam dokumen 2, 5, 7, dan 9, sehingga apabila kita mencari kata *elephant* maka kita langsung bisa menemukan dokumen-dokumen tersebut tanpa harus membuka satu per satu semua dokumen yang ada.

## 2.4 Pendeteksian Bahasa

Pendeteksian bahasa adalah fitur yang terdapat di dalam sistem ini. Sistem didesain untuk mendeteksi dokumen dalam bahasa Inggris dan Indonesia, oleh karena itu dibutuhkan database kata-kata bahasa Inggris dan Indonesia yang telah dibuat terlebih dahulu. Proses yang dilakukan adalah sebagai berikut:

1. Dalam proses indexing sebuah dokumen, tiap-tiap kata dibaca dari dokumen tersebut.
2. Tiap-tiap kata dibandingkan dengan database masing-masing bahasa, apabila cocok maka *score* untuk bahasa tersebut akan ditambah 1 poin.
3. Hasil pendeteksian bahasa adalah bahasa yang mempunyai *score* terbanyak.

## 2.5 Noise Word Filtering

Sistem ini juga menerapkan noise word filtering untuk menghilangkan kata-kata yang dianggap tidak perlu diindeks. Ini dilakukan selain untuk mempercepat performansi juga untuk mengurangi ukuran database. Untuk itu, dibutuhkan database *stop word* dalam bahasa Inggris dan bahasa Indonesia.

Proses noise word filtering adalah sebagai berikut:

1. Deteksi bahasa dilakukan terlebih dahulu. Database *stop word* yang digunakan sesuai dengan bahasa yang dideteksi. Apabila bahasa tidak terdeteksi maka semua database *stop word* digunakan karena bahasa dianggap netral.
2. Setiap kata-kata yang terdapat dalam database *stop word* dieliminasi dan tidak diikutsertakan dalam proses *indexing*.
3. Setiap kata yang muncul kurang dari 2 kali dalam sebuah dokumen juga dieliminasi dan tidak diikutsertakan dalam proses *indexing*.

## 2.6 Fenomena-fenomena Anomali

Dalam sistem ini, ada kemungkinan terjadinya anomali pada data yang diproses. Bentuk-bentuk anomali yang mungkin terjadi antara lain:

1. Bentuk-bentuk anomali yang diakibatkan oleh pengguna.

a. Pengguna yang sengaja memasukkan data yang salah, misalnya memasukkan sebuah dokumen ke tag/kategori yang salah.

b. Pengguna yang tidak sengaja atau keliru memasukkan data, sehingga mengakibatkan sebuah dokumen masuk ke tag/kategori yang salah.

c. Pengguna yang mencoba melakukan perusakan terhadap sistem.

2. Bentuk-bentuk anomali yang diakibatkan oleh kelemahan *folksonomy*.

   a. Satu *tag* dapat mengandung beberapa arti, tergantung konteks. Misalnya, tag *apple* bisa digunakan untuk menyebut buah apel tapi bisa juga digunakan untuk perusahaan Apple Computing.

   b. Satu arti yang sama bisa mempunyai beberapa *tag* yang berbeda. Contohnya *mac* dan *macintosh* mempunyai arti yang sama, tapi sistem tetap membedakan kedua tag tersebut.

   c. Satu kata memiliki ejaan yang berbeda-beda. Contohnya jogja, yogya, dan yogyakarta.

   d. Batasan lingkup sebuah tag tidak jelas. Apabila sebuah bookmark termasuk dalam *tag PHP*, bisa dipastikan dokumen tersebut termasuk dalam *tag programming*, apalagi *computing*. Namun sangat tidak praktis untuk mencantumkan semua tag yang cocok dengan bookmark tersebut karena jumlahnya sangat banyak.

3. Bentuk-bentuk anomali yang disebabkan oleh metode Bayes.

   a. Sebuah dokumen dapat dikategorisasikan secara otomatis ke tag/kategori yang salah.

   b. Sebuah dokumen semestinya masuk ke sebuah tag/kategori tertentu, tetapi berdasarkan kategorisasi otomatis tidak memenuhi syarat.

# Bab III
# Analisis dan Perancangan Sistem

## 3.1 *Gambaran Umum Sistem*

Sistem ini dibuat menggunakan tool PHP versi 5, didukung oleh web server Apache versi 2, dan MySQL versi 3.23.

## 3.2 *Kebutuhan Sistem*

### 3.2.1 Spesifikasi Perangkat Keras

Perangkat keras yang digunakan untuk membuat sistem ini adalah dengan spesifikasi:

- AMD Sempron 2600+
- RAM 512 MB
- Hard disk 120 GB
- Kartu grafis NVidia GeForce FX 5200
- Monitor LG Studioworks E700B
- Keyboard dan mouse

### 3.2.2 Spesifikasi Perangkat Lunak

Perangkat lunak yang dibutuhkan untuk menjalankan sistem ini adalah:

- Microsoft Windows XP Professional Service Pack 2
- Apache versi 2.0.53
- PHP versi 5.0.4
- MySQL versi 3.23.54

## 3.3 *Desain Sistem*

### 3.3.1 Diagram Use Case

Use case diagram menjelaskan manfaat sistem jika dilihat menurut pandangan orang yang berada diluar sistem (*actor*). Diagram ini menunjukkan fungsionalitas suatu sistem atau kelas dan bagaimana sistem berinteraksi dengan dunia luar. Use case untuk sistem ini dapat dilihat sebagai berikut.

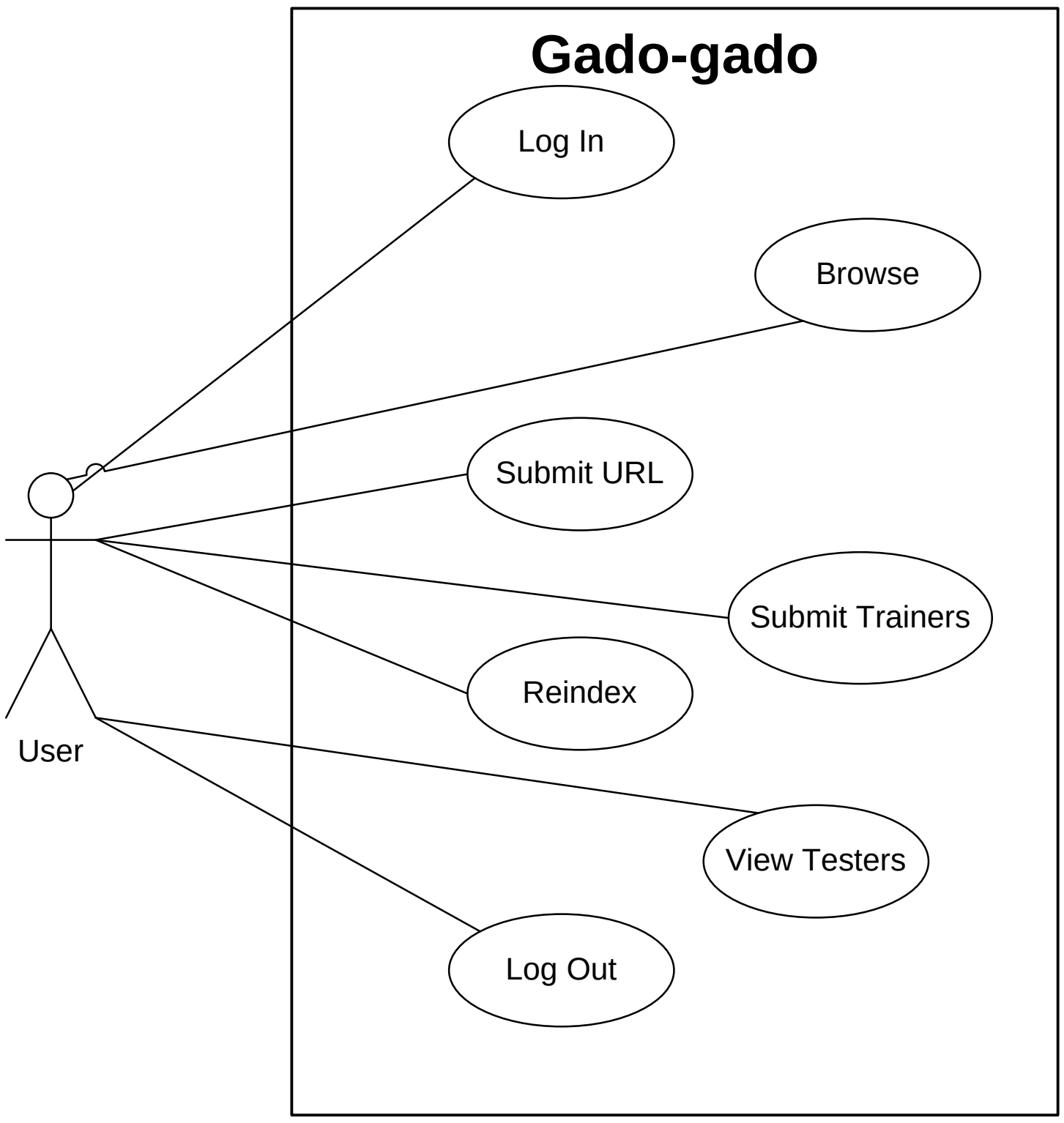

## 3.3.2 Penjelasan Use Case

Penjelasan untuk masing-masing use case adalah sebagai berikut:

### 3.3.2.1 Log In

| Sistem | User |
|---|---|
| 1. Menampilkan halaman log in, yang terdiri dari field username, password, dan tombol Log in. | |
| | 2. Mengisi username dan password.<br>3. Menekan tombol Log in. |
| 4. Mencari username yang dimaksud di dalam database. Apabila username tidak ditemukan, tampilkan pesan kesalahan.<br>5. Mencocokkan password di database dengan password yang diinputkan user. Apabila password tidak cocok, tampilkan pesan kesalahan.<br>6. Set session/cookie untuk menandai bahwa user telah aktif.<br>7. Tampilkan halaman menu utama. | |

### 3.3.2.2 Browse

| Sistem | User |
|---|---|
| 1. Menampilkan daftar tag yang terdapat dalam database. | |
| | 2. Meng-klik salah satu tag yang diinginkan. |
| 3. Menampilkan daftar dokumen sesuai dengan tag yang dipilih oleh user. | |

#### 3.3.2.3 Submit URL

| Sistem | User |
|---|---|
| 1. Menampilkan formulir Submit URL, yang terdiri atas URL dokumen dan daftar tag yang diinginkan. | |
| | 2. Mengisi URL dokumen dan daftar tag yang diinginkan.<br>3. Meng-klik tombol Submit. |
| 4. Mengecek validitas input yang dimasukkan.<br>5. Menambahkan record di database untuk URL dokumen yang dimasukkan. Apabila di database telah ada dokumen dengan URL yang sama, gunakan record tersebut.<br>6. Untuk setiap tag yang diinginkan, daftarkan dokumen pada tag tersebut dalam database.<br>7. Tampilkan pesan sukses. | |

#### 3.3.2.4 Submit Trainers

| Sistem | User |
|---|---|
| 1. Menampilkan halaman menu utama. | |
| | 2. Meng-klik menu “Submit Trainers”. |
| 3. Membuka setiap file dalam folder trainer. Untuk setiap dokumen:<br>a. Menambahkan record di database untuk URL dokumen yang dimasukkan. Apabila di database telah ada dokumen dengan URL yang sama, gunakan record tersebut.<br>b. Untuk setiap tag yang diinginkan, daftarkan dokumen pada tag tersebut dalam database.<br>c. Tampilkan status proses pada layar. | |

#### 3.3.2.5 Reindex

| Sistem | User |
|---|---|
| 1. Menampilkan halaman menu utama. | |
| | 2. Meng-klik menu “Reindex”. |
| 3. Memproses setiap dokumen di database yang belum diindex. Untuk setiap dokumen, lakukan proses sebagai berikut:<br>a. Membuka dokumen tersebut dan mengambil isinya.<br>b. Melakukan ekstraksi kata-kata.<br>c. Melakukan deteksi bahasa dan noise word filtering.<br>d. Melakukan indexing terhadap kata-kata di dalam dokumen tersebut.<br>e. Tampilkan status proses pada layar.<br>4. Untuk setiap *submission queue* dalam database, lakukan proses sebagai berikut: | |

| Sistem | User |
|---|---|
| a. Daftarkan dokumen tersebut pada tag yang dimaksud.<br>b. Tampilkan status proses pada layar. | |

**3.3.2.6 View Testers**

| Sistem | User |
|---|---|
| 1. Menampilkan halaman menu utama.<br><br>3. Membuka setiap file dalam folder tester. Untuk setiap dokumen:<br>a. Lakukan kategorisasi otomatis menggunakan metode Bayes.<br>4. Menampilkan hasil kategorisasi otomatis dalam bentuk tabel. | 2. Meng-klik menu “View Testers”. |

**3.3.2.7 Log Out**

| Sistem | User |
|---|---|
| 1. Menampilkan halaman menu utama.<br><br>3. Me-reset session/cookie untuk menandakan user tidak aktif.<br>4. Menampilkan halaman log in. | 2. Meng-klik menu “Log out”. |

### 3.3.3 Diagram Aktivitas

Diagram aktivitas untuk masing-masing fungsionalitas adalah sebagai berikut:

### 3.3.3.1 Log In

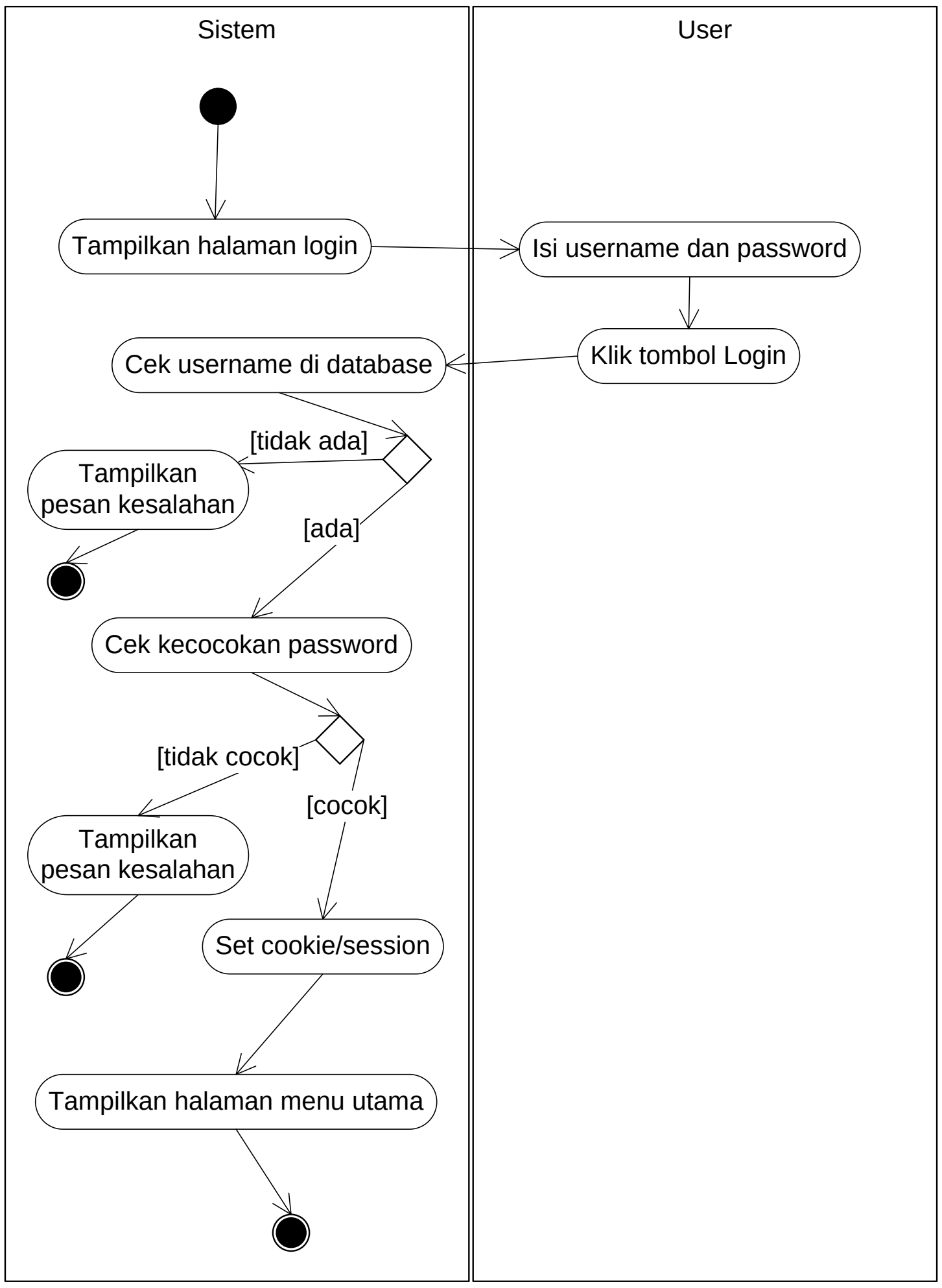

### 3.3.3.2 Browse

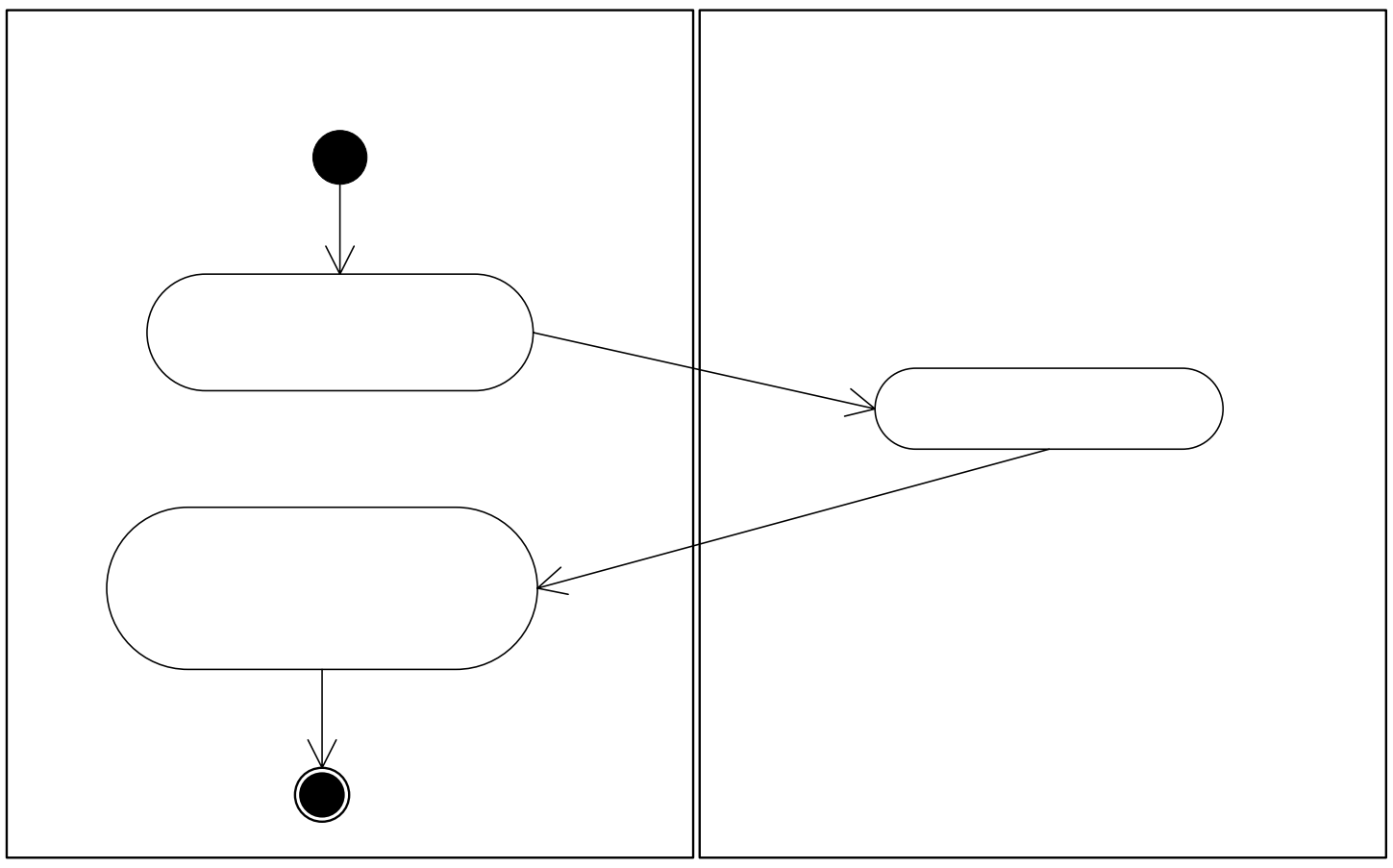

### 3.3.3.3 Submit URL

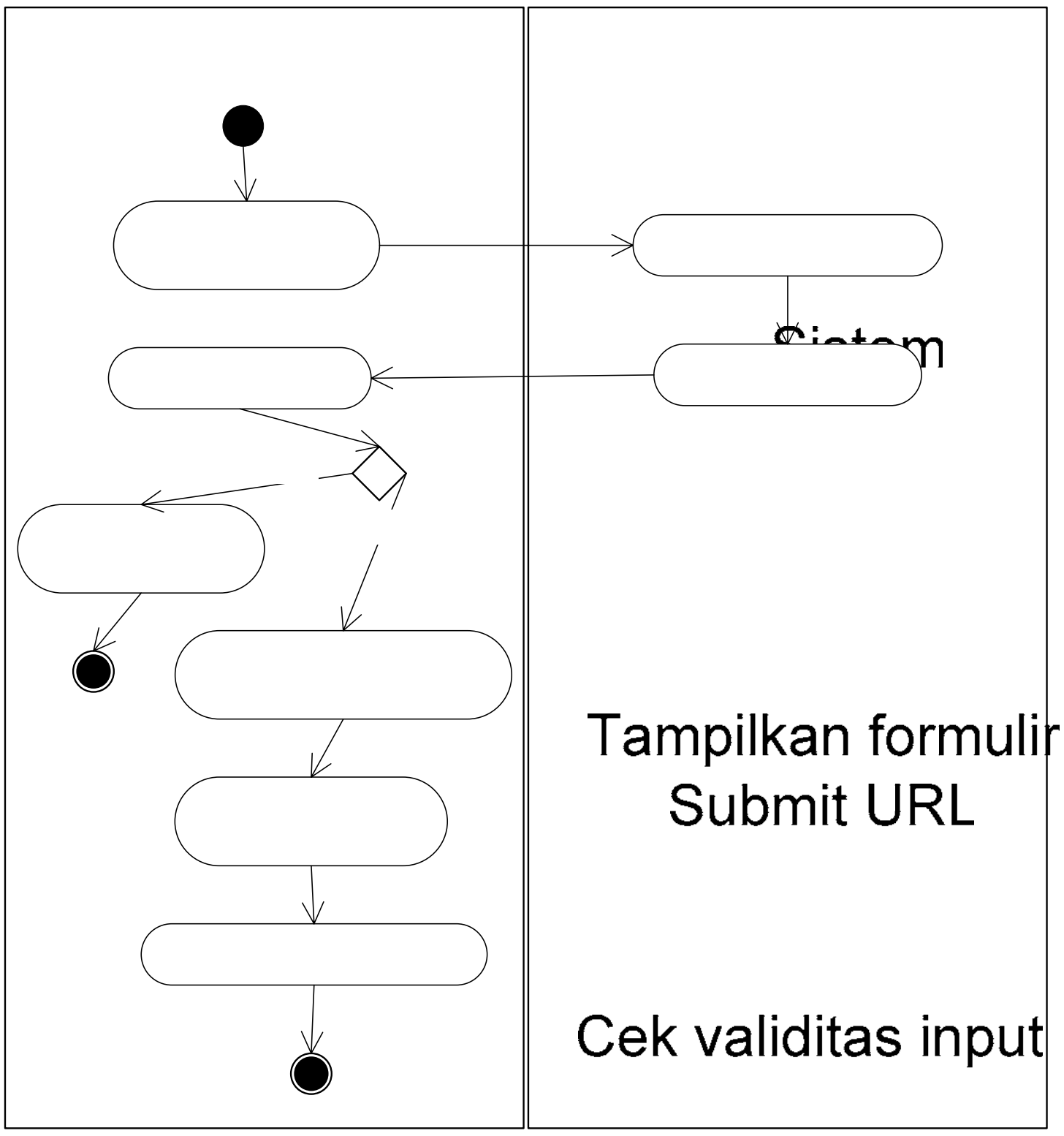

### 3.3.3.4 Submit Trainers

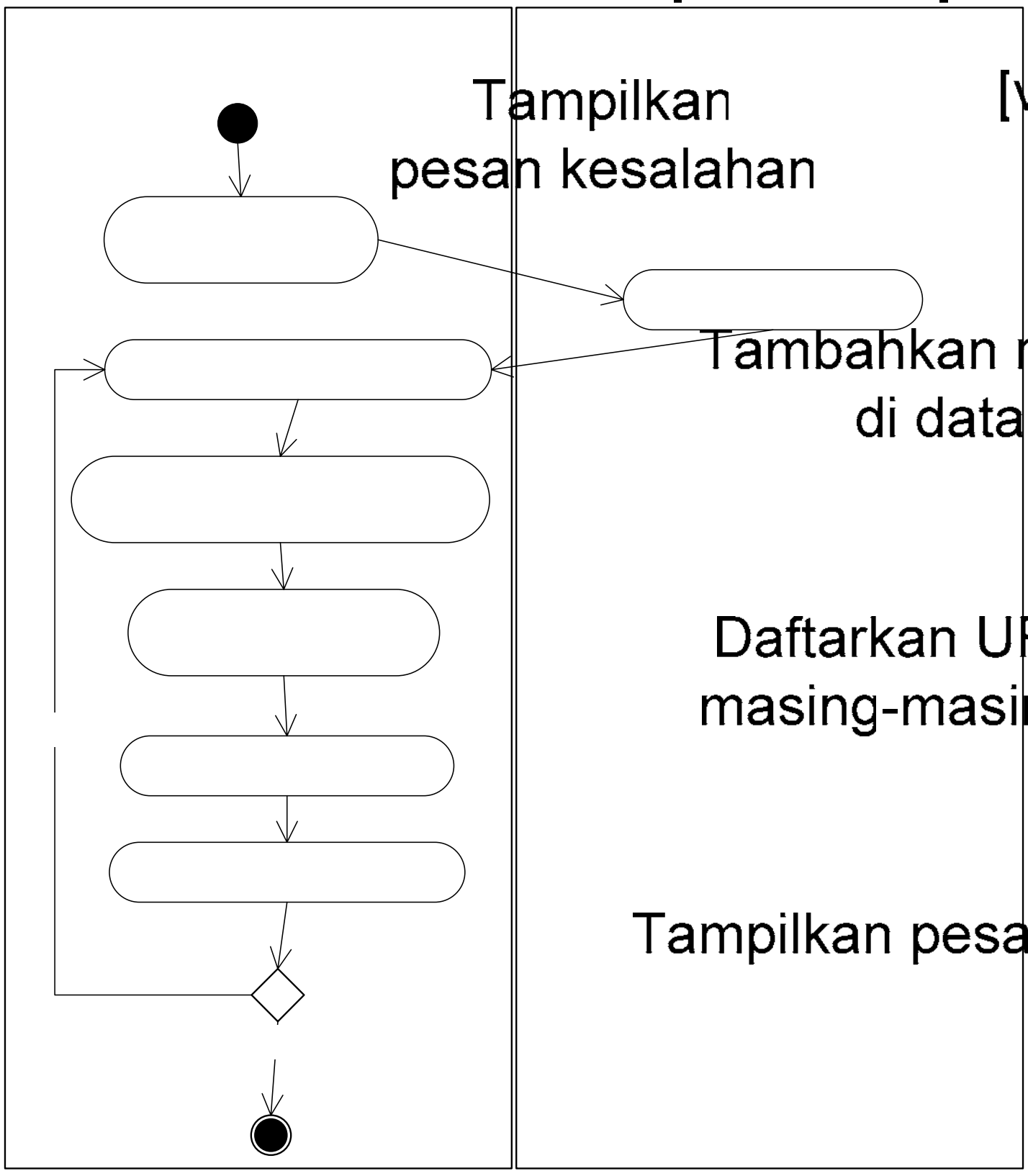

### 3.3.3.5 Reindex

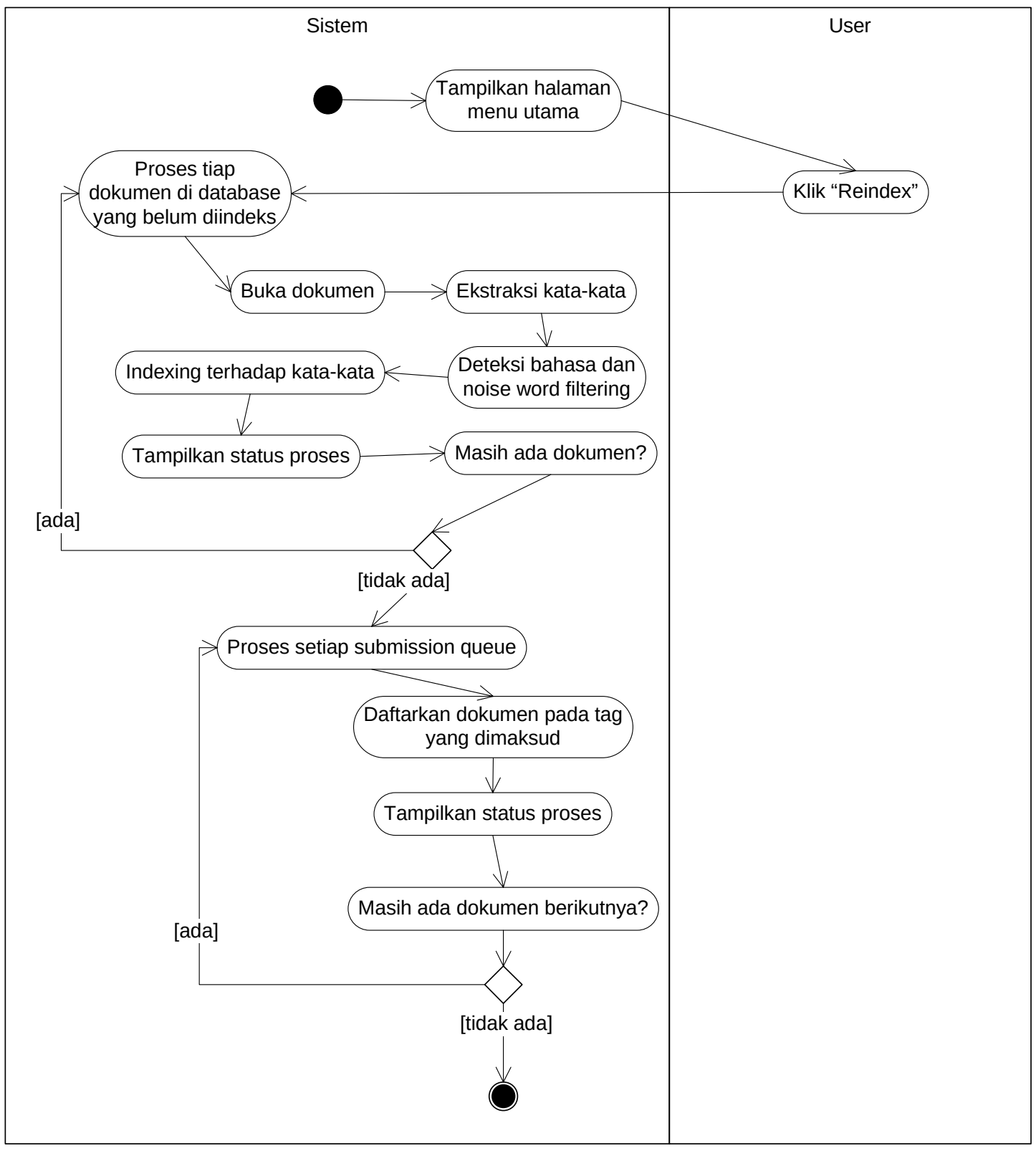

### 3.3.3.6 View Testers

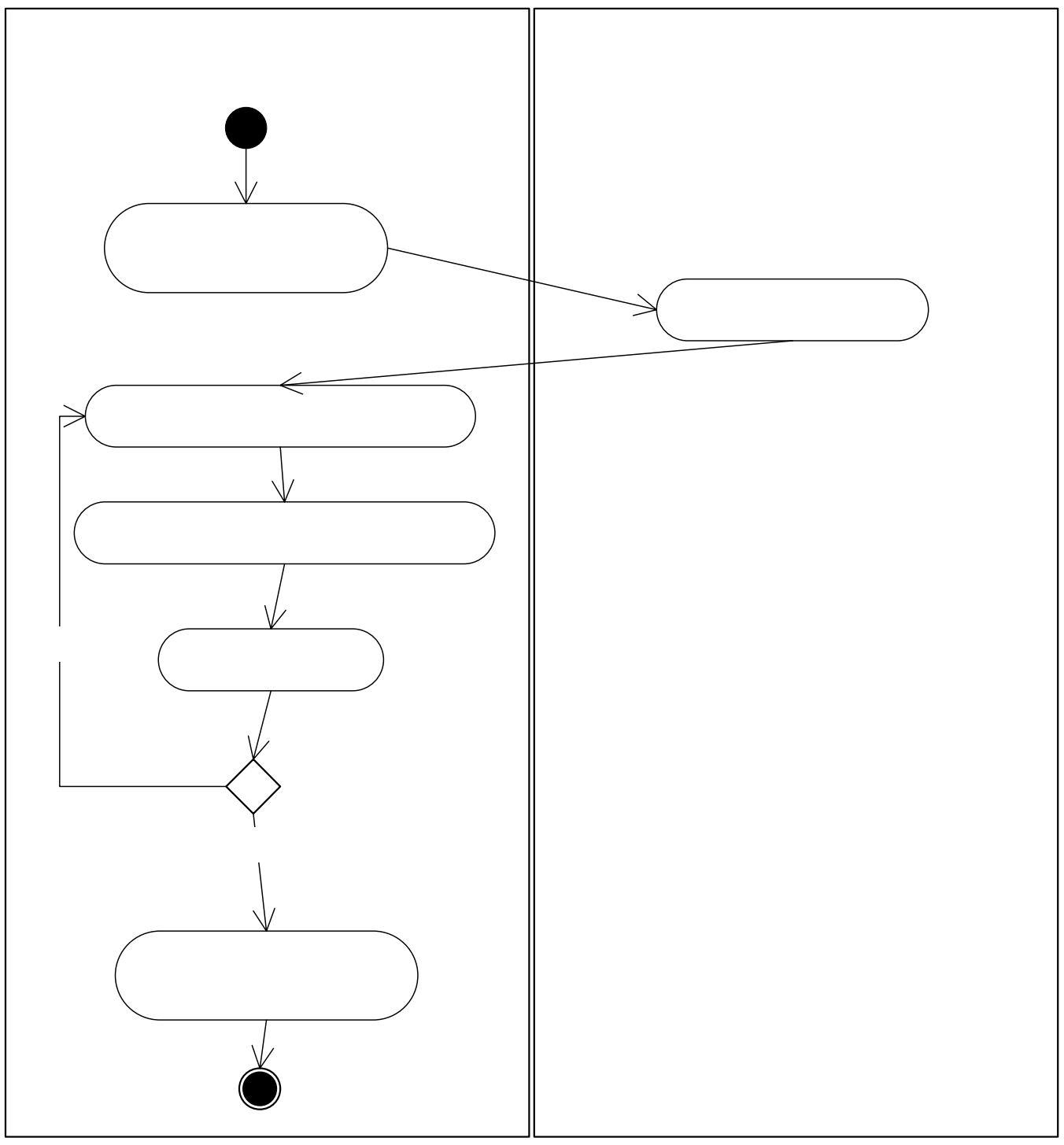

Sistem User

**3.3.3.7 Log Out**

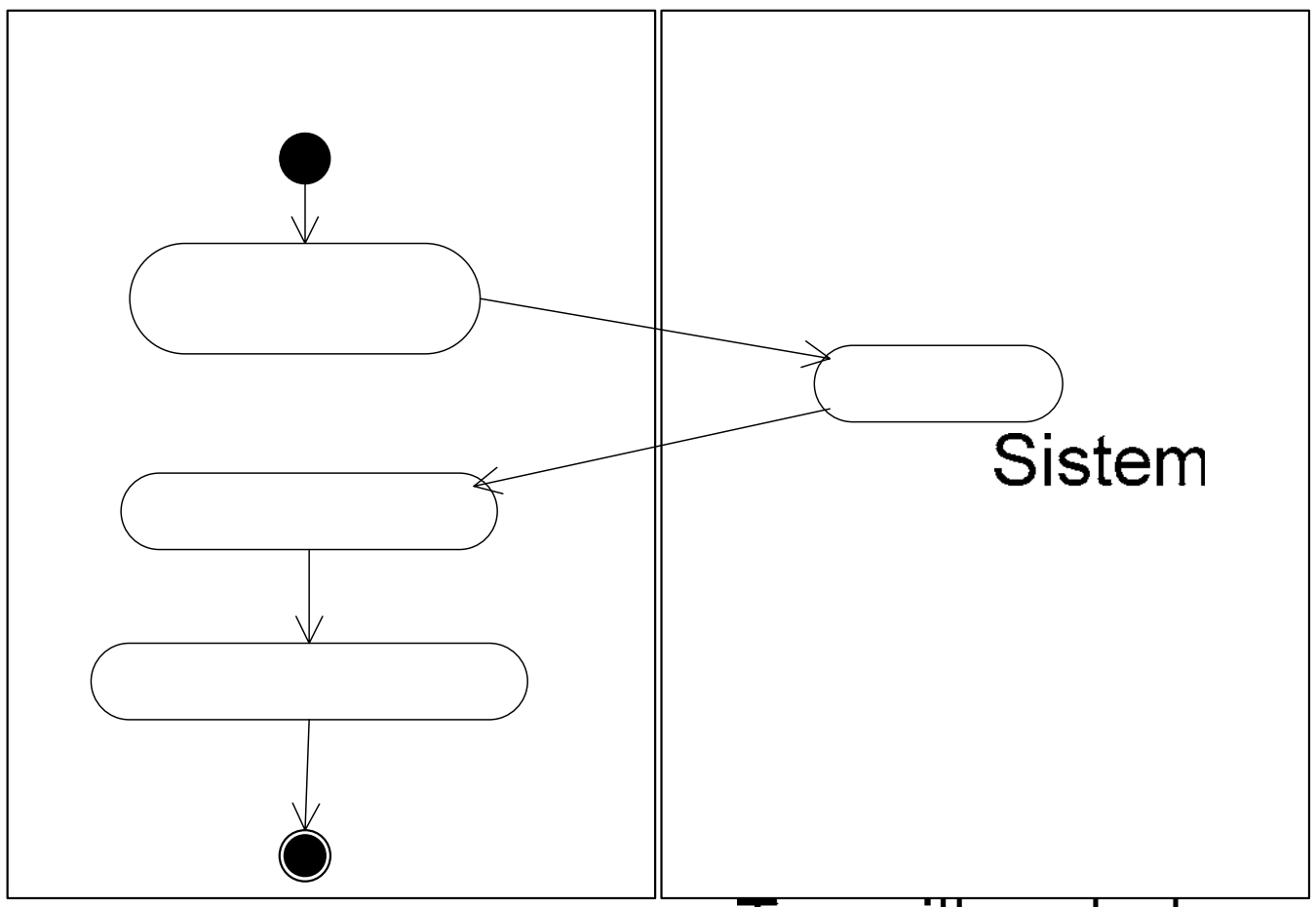

### 3.3.4 Model Konseptual

Model konseptual adalah representasi konsep domain masalah. Dalam model konseptual dijumpai konsep-konsep yang ada pada sistem, asosiasi antar konsep, dan atribut yang dimiliki oleh konsep-konsep tersebut. Model konseptual untuk sistem ini adalah sebagai berikut :

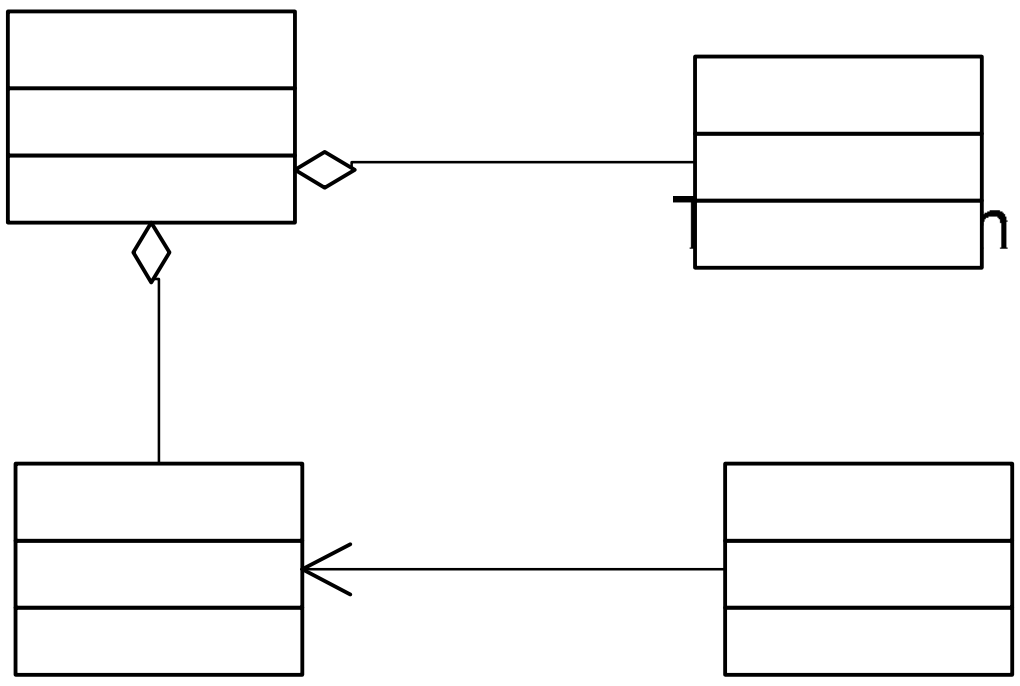

### 3.3.5 Diagram Kelas

Setiap obyek merupakan *instance* dari suatu kelas, dimana kelas tersebut menggambarkan *properties* dan *behaviour* dari setiap jenis obyek. Sebuah diagram kelas menggambarkan kelas-kelas yang terdapat pada sistem dan hubungannya dengan kelas lainnya. Diagram kelas sistem ini adalah sebagai berikut :

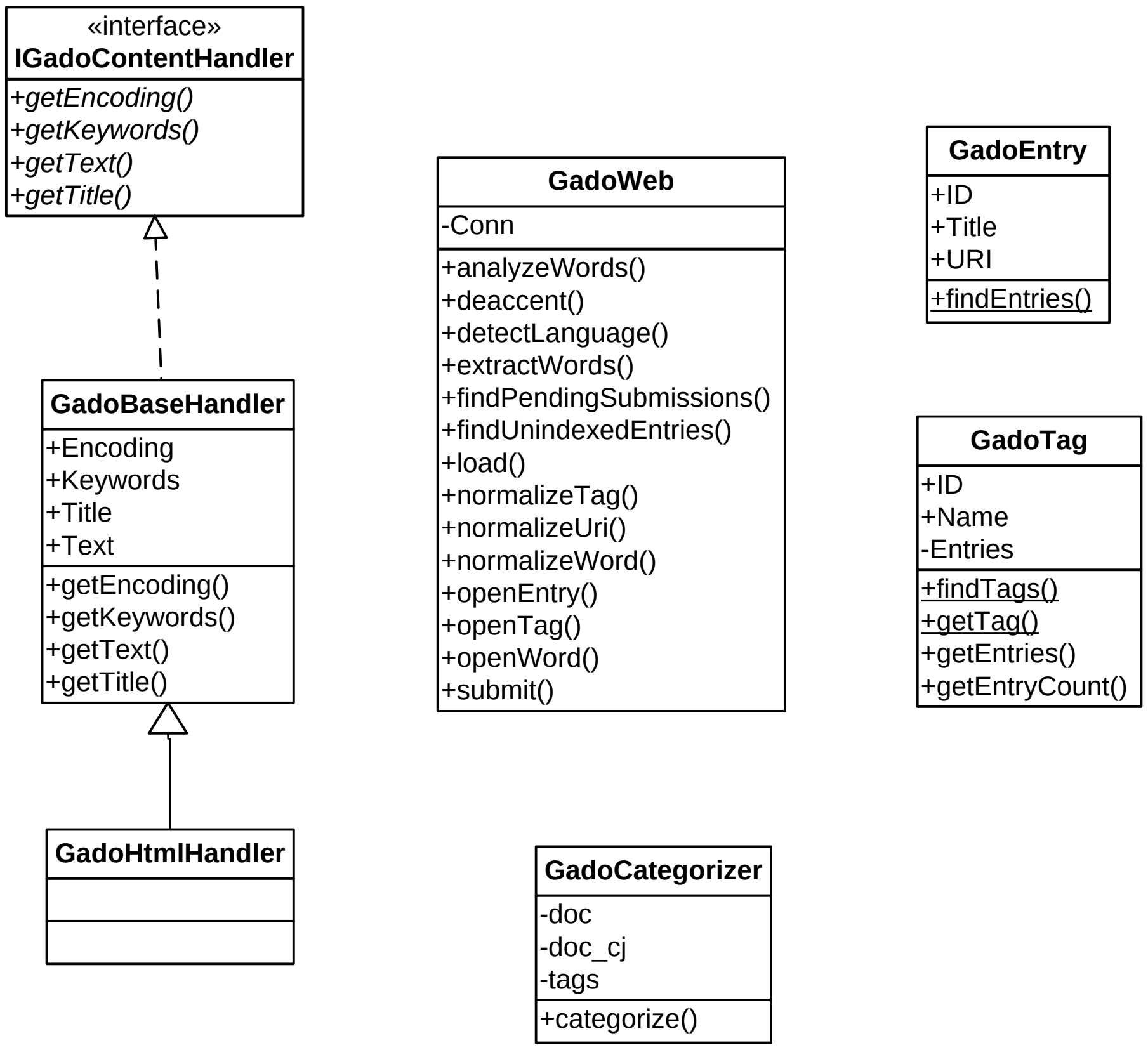

### 3.3.6 Data Model

Data model merupakan penggambaran struktur data yang disimpan di dalam database, dalam kasus ini database yang dipakai adalah MySQL.

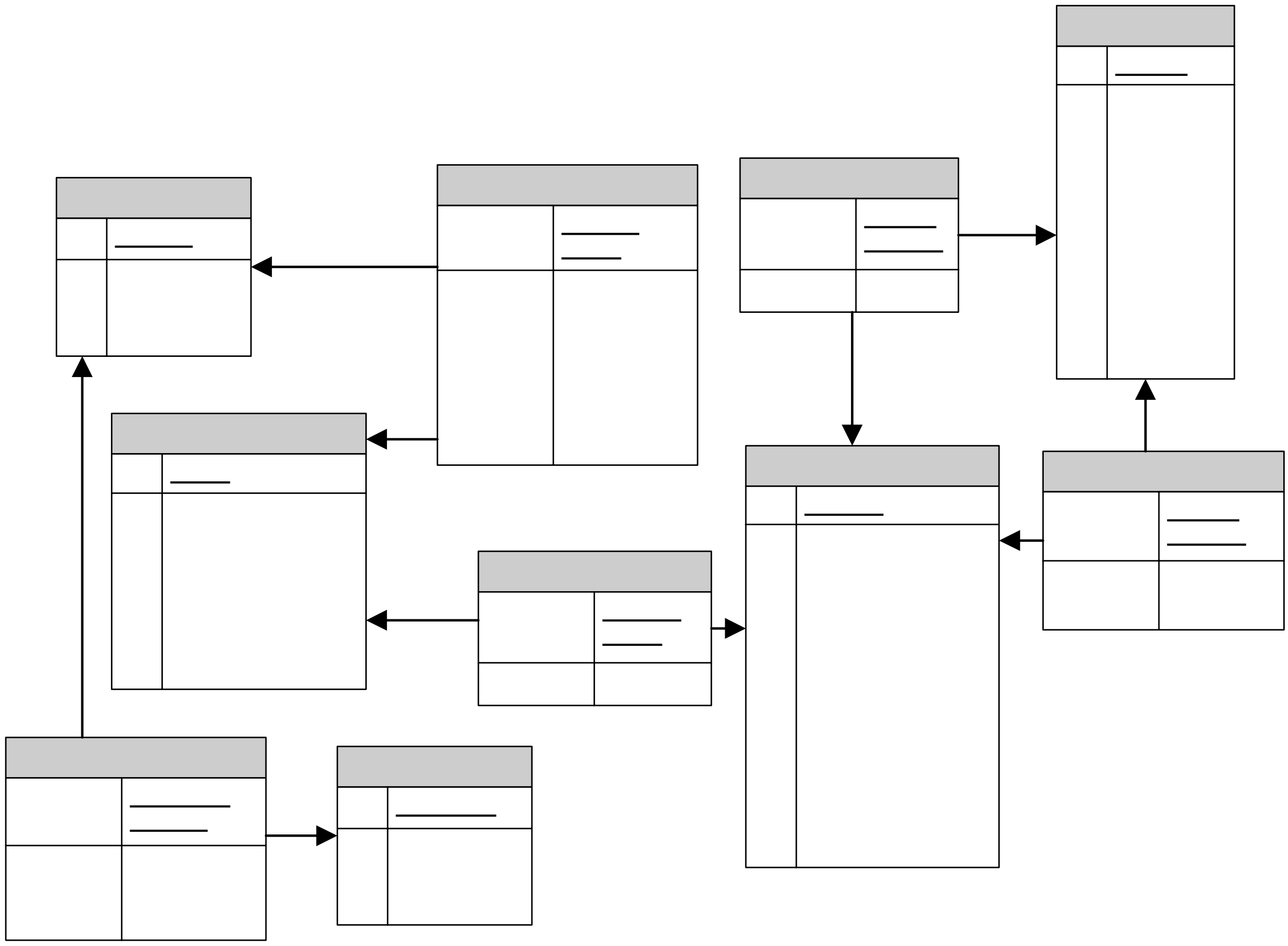

# Bab IV
# Implementasi dan Pengujian

## 4.1 Skenario Pengujian

### 4.1.1 Pengujian Implementasi

Untuk menguji implementasi, kasus uji yang digunakan adalah:

- Sebuah file *trainer* yang dikategorisasikan secara manual ke tag **biz**. Isi file adalah sebagai berikut:

  ```
  money money
  ```

- Sebuah file *trainer* yang dikategorisasikan secara manual ke tag **net**. Isi file adalah sebagai berikut:

  ```
  cable cable
  ```

- Sebuah file *tester*. Isi file adalah sebagai berikut:

  ```
  cable cable cable money money
  ```

- File *tester* tersebut berisi kata-kata dari kedua buah tag, namun jumlah kata cable (untuk tag "net") lebih banyak. Oleh karena itu, keluaran yang diharapkan adalah sistem dapat mengkategorisasikan file *tester* tersebut ke dua tag, namun dengan nilai probabilitas lebih tinggi untuk tag "net" daripada tag "biz".

Diketahui:

d : sekumpulan kata-kata dalam dokumen tester = {cable, cable, cable, money, money}

|V| : himpunan kata-kata yang diketahui = {cable, money}

Berdasarkan hasil training, *prior probability* untuk masing-masing tag adalah sebagai berikut:

$$p(c_{net}) = 0{,}5$$

$$p(c_{biz}) = 0{,}5$$

Pertama kita harus menghitung probabilitas masing-masing kata terhadap masing-masing kategori, dengan menggunakan *smoothing* Laplace untuk menghindari nilai 0.

**Perhitungan untuk tag "net":**

$$p(w_{cable} \mid c_{net}) = \frac{1 + \#words(c_{net}, w_{cable})}{|V| + \#words(c_{net})} = \frac{1+2}{2+2} = \frac{3}{4} = 0{,}75$$

$$p\left(w_{money} \mid c_{net}\right) = \frac{1 + \#words(c_{net}, w_{money})}{|V| + \#words\left(c_{net}\right)} = \frac{1+0}{2+2} = \frac{1}{4} = 0{,}25$$

$$p\left(w_{cable} \mid \overline{c_{net}}\right) = \frac{1 + \#words(\overline{c_{net}}, w_{cable})}{|V| + \#words\left(\overline{c_{net}}\right)} = \frac{1+0}{2+2} = \frac{1}{4} = 0{,}25$$

$$p\left(w_{money} \mid \overline{c_{net}}\right) = \frac{1 + \#words(\overline{c_{net}}, w_{money})}{|V| + \#words\left(\overline{c_{net}}\right)} = \frac{1+2}{2+2} = \frac{3}{4} = 0{,}75$$

Dari nilai-nilai di atas kita dapat menghitung probabilitas dokumen terhadap tag “net”:

$$p\left(d \mid c_{net}\right) = \frac{0{,}75^3}{3!} \times \frac{0{,}25^2}{2!} = 0{,}0021972$$

$$p\left(d \mid \overline{c_{net}}\right) = \frac{0{,}25^3}{3!} \times \frac{0{,}75^2}{2!} = 0{,}00073242$$

Lalu kita dapat menghitung probabilitas dokumen tersebut:

$$\begin{aligned} p(d) &= p\left(d \mid c_{net}\right) \times p\left(c_{net}\right) + p\left(d \mid \overline{c_{net}}\right) \times p\left(\overline{c_{net}}\right) \\ &= 0{,}0021972 \times 0{,}5 + 0{,}00073242 \times 0{,}5 = 0{,}0010986 + 0{,}00036621 \\ &= 0{,}0014648 \end{aligned}$$

Setelah itu kita dapat mengetahui berapa probabilitas tag tersebut:

$$p\left(c_{net} \mid d\right) = \frac{p\left(d \mid c_{net}\right) \times p\left(c_{net}\right)}{p(d)} = \frac{0{,}0021972 \times 0{,}5}{0{,}0014648} = 0{,}75$$

Berarti probabilitas untuk tag “net” terhadap dokumen tester tersebut adalah 75%.

**Perhitungan untuk tag “biz”:**

$$p\left(w_{cable} \mid c_{biz}\right) = \frac{1 + \#words(c_{biz}, w_{cable})}{|V| + \#words\left(c_{biz}\right)} = \frac{1+0}{2+2} = \frac{1}{4} = 0{,}25$$

$$p\left(w_{money} \mid c_{biz}\right) = \frac{1 + \#words(c_{biz}, w_{money})}{|V| + \#words\left(c_{biz}\right)} = \frac{1+2}{2+2} = \frac{3}{4} = 0{,}75$$

$$p\left(w_{cable} \mid \overline{c_{biz}}\right) = \frac{1 + \#words(\overline{c_{biz}}, w_{cable})}{|V| + \#words\left(\overline{c_{biz}}\right)} = \frac{1+2}{2+2} = \frac{3}{4} = 0{,}75$$

$$p\left(w_{money} \mid \overline{c_{biz}}\right) = \frac{1 + \#words(\overline{c_{biz}}, w_{money})}{|V| + \#words\left(\overline{c_{biz}}\right)} = \frac{1+0}{2+2} = \frac{1}{4} = 0{,}25$$

Dari nilai-nilai di atas kita dapat menghitung probabilitas dokumen terhadap tag “biz”:

$$p(d \mid c_{biz}) = \frac{0{,}25^3}{3!} \times \frac{0{,}75^2}{2!} = 0{,}00073242$$

$$p(d \mid \overline{c_{biz}}) = \frac{0{,}75^3}{3!} \times \frac{0{,}25^2}{2!} = 0{,}0021972$$

Lalu kita dapat menghitung probabilitas dokumen tersebut (nilai ini seharusnya sama dengan perhitungan sebelumnya untuk tag “net”):

$$\begin{aligned} p(d) &= p(d \mid c_{biz}) \times p(c_{biz}) + p(d \mid \overline{c_{biz}}) \times p(\overline{c_{biz}}) \\ &= 0{,}00073242 \times 0{,}5 + 0{,}0021972 \times 0{,}5 = 0{,}00036621 + 0{,}0010986 \\ &= 0{,}0014648 \end{aligned}$$

Setelah itu kita dapat mengetahui berapa probabilitas tag tersebut:

$$p(c_{biz} \mid d) = \frac{p(d \mid c_{biz}) \times p(c_{biz})}{p(d)} = \frac{0{,}00073242 \times 0{,}5}{0{,}0014648} = 0{,}25$$

Berarti probabilitas untuk tag “biz” terhadap dokumen tester tersebut adalah 25%.

Dari perhitungan untuk dua tag di atas kita dapat menyimpulkan bahwa hasil kategorisasi untuk dokumen tester tersebut adalah:

- Untuk tag “net” = 75%
- Untuk tag “biz” = 25%

Hasil output sistem adalah sebagai berikut:

| No. | Tester | Results |
|---|---|---|
| 1. | D:\u1294\gado.gauldong.net\tester\net\test_mixed.txt<br>Expected: **net** | • **net : 75.00%**<br>• biz : 25.00% |

Dengan demikian dapat disimpulkan bahwa hasil perhitungan sistem sesuai dengan perhitungan secara manual.

### 4.1.2 Pengujian Performansi

Pengujian sistem dilakukan dengan cara:

1. Mendaftarkan beberapa dokumen sebagai sampel *trainer*.
    a. Melakukan analisa performansi pada pendaftaran dokumen.
2. Melakukan indexing.
    a. Melakukan analisa performansi pada indexing.

### 4.1.3 Pengujian Akurasi Kategorisasi Otomatis

Pengujian sistem dilakukan dengan cara:

1. Mengumpulkan sampel *tester*.
    a. Setiap sampel ditandai dengan tag hasil kategorisasi secara manual.
2. Melakukan kategorisasi otomatis pada sampel *tester*.
3. Melakukan analisa akurasi pada kategorisasi otomatis. Kriteria akurasi sebagai berikut:
    a. Apabila seluruh kategori manual yang diharapkan dapat diberikan oleh kategorisasi otomatis dengan probabilitas 50% - 100%, maka sampel tester tersebut dinyatakan berhasil baik.
    b. Apabila tidak seluruh kategori manual yang diharapkan dapat diberikan oleh kategorisasi otomatis atau mendapatkan probabilitas di bawah 50%, maka sampel tester tersebut dinyatakan berhasil namun tidak meyakinkan.
    c. Apabila kategori manual tidak dapat diberikan oleh kategorisasi otomatis, maka sampel tester tersebut dinyatakan gagal.

### 4.1.4 Alternatif Pengujian Akurasi

Sebagai bahan pertimbangan, pengujian akurasi juga dapat dilakukan dengan cara sebagai berikut:

1. Sejumlah responden dipilih sebagai penilai dalam pengujian. Latar belakang pendidikan responden disesuaikan dengan pengguna yang akan menggunakan sistem.
2. Sejumlah dokumen tester dipilih sebagai objek pengujian. Materi dokumen tester hendaknya disesuaikan dengan latar belakang pendidikan responden.
3. Kategorisasi otomatis terhadap sejumlah dokumen tester dilakukan.
4. Para responden dipersilakan membaca masing-masing dokumen tester dan melakukan penilaian untuk masing-masing dokumen.
    - Untuk setiap dokumen, responden dapat memilih satu atau lebih kategori yang cocok dari daftar kategori yang sudah tersedia dalam database.

- Karena menggunakan sistem klasifikasi folksonomy, responden juga dapat menulis nama kategori sendiri apabila kategori yang diinginkan belum tersedia dalam database.
- Responden dapat melihat hasil kategorisasi otomatis, dan menyatakan setuju atau tidak setuju terhadap hasil kategorisasi otomatis tersebut.

## 4.2 Kasus Uji yang Digunakan

### 4.2.1 Pengujian Performansi

Untuk pengujian performansi submission dan indexing, kasus uji yang digunakan adalah:

**Tabel 3. Kasus uji performansi**

| Tag | Tag tambahan | Kasus uji |
|---|---|---|
| cerpen | tulisan | 4 dokumen |
| cinta | cowokcewek | 17 dokumen |
| cowokcewek | | 10 dokumen |
| english | | 35 dokumen |
| html | computer | 29 dokumen |
| indonesian | | 55 dokumen |
| persahabatan | | 19 dokumen |
| php | computer | 26 dokumen |
| puisi | tulisan | 13 dokumen |
| sex | cinta | 14 dokumen |

### 4.2.2 Pengujian Kategorisasi Otomatis

Untuk pengujian akurasi kategorisasi otomatis, kasus uji yang digunakan adalah:

**Tabel 4. Kasus uji akurasi kategorisasi otomatis**

| Tag | Tag tambahan | Kasus uji |
|---|---|---|
| cerpen | tulisan | 4 dokumen |
| cinta | cowokcewek | 4 dokumen |
| coba_love | | 1 dokumen |
| coba_php | | 1 dokumen |
| cowokcewek | | 4 dokumen |
| english | | 4 dokumen |
| html | computer | 4 dokumen |
| indonesian | | 4 dokumen |
| persahabatan | | 4 dokumen |
| php | computer | 4 dokumen |
| puisi | tulisan | 4 dokumen |
| sex | cinta | 4 dokumen |

Jumlah tester yang digunakan adalah 13 dokumen dengan rincian sebagai berikut:

| **Tester** | **Kategorisasi Manual** |
|---|---|
| D:\u1294\gado.gauldong.net\tester\cerpen,tulisan\2003_10_26_cerpenku_archive.html<br>Tester ini merupakan cerita pendek, oleh karena itu dikategorisasikan secara manual ke tag cerpen dan tulisan. | cerpen<br>tulisan |
| D:\u1294\gado.gauldong.net\tester\cinta,cowokcewek\tips.php_tips=4.html<br>Tester ini berisi tips tentang cinta dan hubungan antara cowok dan cewek. | cinta<br>cowokcewek |
| D:\u1294\gado.gauldong.net\tester\coba_love\coba_love_tester.html<br>Tester ini adalah dokumen sederhana yang digunakan untuk menguji apakah dapat mendeteksi tag coba_love dengan baik. | coba_love |
| D:\u1294\gado.gauldong.net\tester\coba_mixed\coba_mixed_tester.html<br>Tester ini adalah dokumen yang berisi gabungan antara coba_php dan coba_love, untuk menguji apakah sistem dapat mendeteksi kedua tag tersebut. | coba_php<br>coba_love |
| D:\u1294\gado.gauldong.net\tester\coba_php\coba_php_tester.html<br>Tester ini adalah dokumen sederhana yang digunakan untuk menguji apakah dapat mendeteksi tag coba_php dengan baik. | coba_php |
| D:\u1294\gado.gauldong.net\tester\cowokcewek,english\3seconds.shtml.htm<br>Tester ini berisi tips untuk cowok dan cewek dalam bahasa Inggris. | cowokcewek<br>english |
| D:\u1294\gado.gauldong.net\tester\english\about.html<br>Tester ini adalah dokumen dalam bahasa Inggris. | english |
| D:\u1294\gado.gauldong.net\tester\html,computer\about.html<br>Tester ini adalah dokumen spesifikasi HTML dalam bahasa Inggris. | html<br>computer |
| D:\u1294\gado.gauldong.net\tester\indonesian\1000 Burung Kertas.txt<br>Tester ini merupakan artikel dalam bahasa Indonesia. | indonesian |
| D:\u1294\gado.gauldong.net\tester\persahabatan,tulisan\A Good Friend.html<br>Tester ini merupakan tulisan yang berisi tentang persahabatan. | persahabatan<br>tulisan |
| D:\u1294\gado.gauldong.net\tester\php,computer\Exceptions.htm<br>Tester ini merupakan dokumen tentang PHP. | php<br>computer |
| D:\u1294\gado.gauldong.net\tester\puisi,tulisan\Anak Panah.txt<br>Tester ini merupakan puisi. | puisi<br>tulisan |
| D:\u1294\gado.gauldong.net\tester\sex,cinta\6 Langkah Foreplay Sempurna.htm<br>Tester ini merupakan artikel tentang cinta. | sex<br>cinta |

Jumlah responden untuk kategorisasi manual adalah 1 orang yaitu saya sendiri.

## *4.3 Pelaksanaan Uji Coba*

Uji coba dilakukan dengan memperhatikan waktu yang diperlukan untuk melakukan sebuah proses.

### 4.3.1 Pengujian pada Operasi Pendaftaran Dokumen

Pada sistem ini dokumen tidak langsung diproses namun didaftarkan atau dimasukkan terlebih dulu ke sebuah *submission queue*. Pada waktu lain, daftar ini akan diproses dan dokumen-dokumen akan dimasukkan ke dalam index.

Pada pengecekan terakhir, untuk mendaftarkan 193 dokumen membutuhkan waktu 3,63 detik. Sehingga rata-ratanya adalah 0,0188 detik per dokumen atau sekitar 18,8 milidetik per dokumen.

### 4.3.2 Pengujian pada Operasi Indexing Dokumen

Untuk melakukan operasi indexing yaitu membaca tiap-tiap dokumen dan memasukkan isinya ke dalam database dalam bentuk sudah di-index membutuhkan waktu 259,96 detik untuk 193 dokumen atau sekitar 1,34 detik per dokumen.

Setelah operasi indexing, tiap-tiap dokumen juga harus didaftarkan ke tag-nya masing-masing. Operasi ini membutuhkan waktu 554,53 detik untuk ke-193 dokumen tersebut, atau rata-rata sekitar 2,87 detik per dokumen.

Secara total, operasi indexing membutuhkan waktu 814,49 detik atau sekitar 4,22 detik per dokumen.

### 4.3.3 Pengujian pada Operasi Tester

Hasil tester yang didapatkan adalah sebagai berikut:

**Tabel 5. Hasil pengujian tester**

| No. | Tester | Results |
|---|---|---|
| 1. | D:\u1294\gado.gauldong.net\tester\cerpen,tulisan\2003_10_26_cerpenku_archive.html<br>Expected: **cerpen,tulisan** | • **tulisan : 100.00%**<br>• **cerpen : 100.00%**<br><br>Status: berhasil baik |
| 2. | D:\u1294\gado.gauldong.net\tester\cinta,cowokcewek\tips.php_tips=4.html<br>Expected: **cinta,cowokcewek** | • **cowokcewek : 100.00%**<br>• **cinta : 100.00%**<br><br>Status: berhasil baik |
| 3. | D:\u1294\gado.gauldong.net\tester\coba_love\coba_love_tester.html<br>Expected: **coba_love** | • **cowokcewek : 90.50%**<br>• coba_love : 36.41%<br>• cinta : 17.89%<br>• english : 9.82%<br>• persahabatan : 3.31%<br>• indonesian : 3.31%<br>• tulisan : 1.89%<br><br>Status: berhasil tidak meyakinkan |
| 4. | D:\u1294\gado.gauldong.net\tester\coba_mixed\coba_mixed_tester.html<br>Expected: **coba_mixed** | • php : 9.51%<br>• coba_php : 4.34%<br>• computer : 3.72%<br>• coba_love : 2.36%<br><br>Status: berhasil tidak meyakinkan |
| 5. | D:\u1294\gado.gauldong.net\tester\coba_php\coba_php_tester.html<br>Expected: **coba_php** | • **php : 90.34%**<br>• **computer : 85.91%**<br>• **coba_php : 71.47%**<br><br>Status: berhasil baik |

| No. | Tester | Results |
|---|---|---|
| 6. | D:\u1294\gado.gauldong.net\tester\cowokcewek,english\3seconds.shtml.htm<br>Expected: **cowokcewek,english** | • **english : 100.00%**<br>• **cowokcewek : 100.00%**<br><br>Status: berhasil baik |
| 7. | D:\u1294\gado.gauldong.net\tester\english\about.html<br>Expected: **english** | • **computer : 100.00%**<br>• **html : 100.00%**<br><br>Status: gagal |
| 8. | D:\u1294\gado.gauldong.net\tester\html,computer\about.html<br>Expected: **html,computer** | • **computer : 100.00%**<br>• **html : 100.00%**<br><br>Status: berhasil baik |
| 9. | D:\u1294\gado.gauldong.net\tester\indonesian\1000 Burung Kertas.txt<br>Expected: **indonesian** | • **tulisan : 100.00%**<br>• indonesian : 11.30%<br>• persahabatan : 2.30%<br><br>Status: berhasil tidak meyakinkan |
| 10. | D:\u1294\gado.gauldong.net\tester\persahabatan,tulisan\A Good Friend.html<br>Expected: **persahabatan,tulisan** | • **persahabatan : 100.00%**<br>• **tulisan : 100.00%**<br><br>Status: berhasil baik |
| 11. | D:\u1294\gado.gauldong.net\tester\php,computer\Exceptions.htm<br>Expected: **php,computer** | • **computer : 100.00%**<br>• **php : 100.00%**<br><br>Status: berhasil baik |
| 12. | D:\u1294\gado.gauldong.net\tester\puisi,tulisan\Anak Panah.txt<br>Expected: **puisi,tulisan** | • **tulisan : 100.00%**<br>• **puisi : 100.00%**<br>• cerpen : 3.24%<br><br>Status: berhasil baik |
| 13. | D:\u1294\gado.gauldong.net\tester\sex,cinta\6 Langkah Foreplay Sempurna.htm<br>Expected: **sex,cinta** | • **sex : 100.00%**<br>• **cinta : 100.00%**<br><br>Status: berhasil baik |

Penggunaan *smoothing* Laplace dapat mengakibatkan munculnya kategori-kategori yang tidak relevan meski prosentasenya sangat kecil, untuk itu sistem memfilternya dengan menggunakan *threshold* 1,0% (hasil di bawah 1,0% diabaikan).

Dengan demikian dapat diambil kesimpulan:

- 9 dari 13 (69,23%) tester berhasil baik
- 3 dari 13 (23,08%) tester berhasil namun tidak meyakinkan
- 1 dari 13 (7,69%) tester dinyatakan gagal

## 4.4 *Ringkasan Hasil Pengujian*

Hasil uji coba pada sistem kami di atas dirangkum pada tabel di bawah ini:

**Tabel 6. Hasil pengujian performansi**

| No. | Pengujian | Lama waktu | |
|---|---|---|---|
| | | Total | Per Dokumen |
| 1. | Pendaftaran Dokumen | 3,63 s | 0,0188 s |
| 2. | Indexing (Tahap 1) | 259,96 s | 1,34 s |
| 3. | Indexing (Tahap 2) | 554,53 s | 2,87 s |
| 4. | Indexing (Total) | 814,49 s | 4,22 s |

Pengujian Klasifikasi:

**Tabel 7. Hasil pengujian klasifikasi**

| No. | Pengujian | Hasil | |
|---|---|---|---|
| | | Jumlah tester | Prosentase |
| 1. | Berhasil baik | 9 dari 13 | 69,23% |
| 2. | Berhasil tidak meyakinkan | 3 dari 13 | 23,08% |
| 3. | Gagal | 1 dari 13 | 7,69% |

# Bab V
# Kesimpulan dan Saran

## 5.1 Kesimpulan

Dari hasil pembangunan sistem ini serta dari hasil uji coba yang telah dilakukan, dapat ditarik beberapa kesimpulan sebagai berikut:

1. Metode multinomial naive Bayes yang dipakai oleh Gado-gado dapat digunakan sebagai metode klasifikasi atau kategorisasi teks dengan akurasi hampir 70% (sesuai dengan hasil pengujian pada Bab IV).
2. Algoritma multinomial naive Bayes yang dipakai oleh Gado-gado memiliki tingkat kegagalan sekitar 7%.
3. Akurasi algoritma tersebut berlaku untuk penggunaan 15 tag (kategori).
4. Kategorisasi otomatis yang dipakai bisa digunakan untuk mendukung atau sebagai sistem pemberi saran bagi kategorisasi manual. Sejumlah responden dibutuhkan untuk menilai apakah kategorisasi otomatis yang dihasilkan telah sesuai dengan yang diinginkan, karena hasil kategorisasi bersifat subjektif.
5. Pada operasi indexing, rata-rata lama waktu pemrosesan adalah 4,22 detik per dokumen.

## 5.2 Saran

Dari hasil pembangunan sistem ini serta dari hasil uji coba yang telah dilakukan, saran-saran yang dapat diberikan antara lain sebagai berikut:

1. Dalam menggunakan metode multinomial naive Bayes, perlu dilakukan atau diadakan teknik yang menambahkan fungsionalitas kategorisasi selain dengan multinomial naive Bayes apalagi metode ini gagal melakukan fungsinya.
2. Metode multinomial naive Bayes baik digunakan untuk mendukung kategorisasi teks secara manual sebagai fungsionalitas pendukung.
3. Waktu indexing yang lama membutuhkan konfigurasi sistem hardware yang cukup tinggi agar dapat meningkatkan kinerja atau dapat ditambahkan optimalisasi dari segi perangkat lunak.

# Daftar Pustaka

# Lampiran I
# Print Screen Aplikasi

## Halaman Login

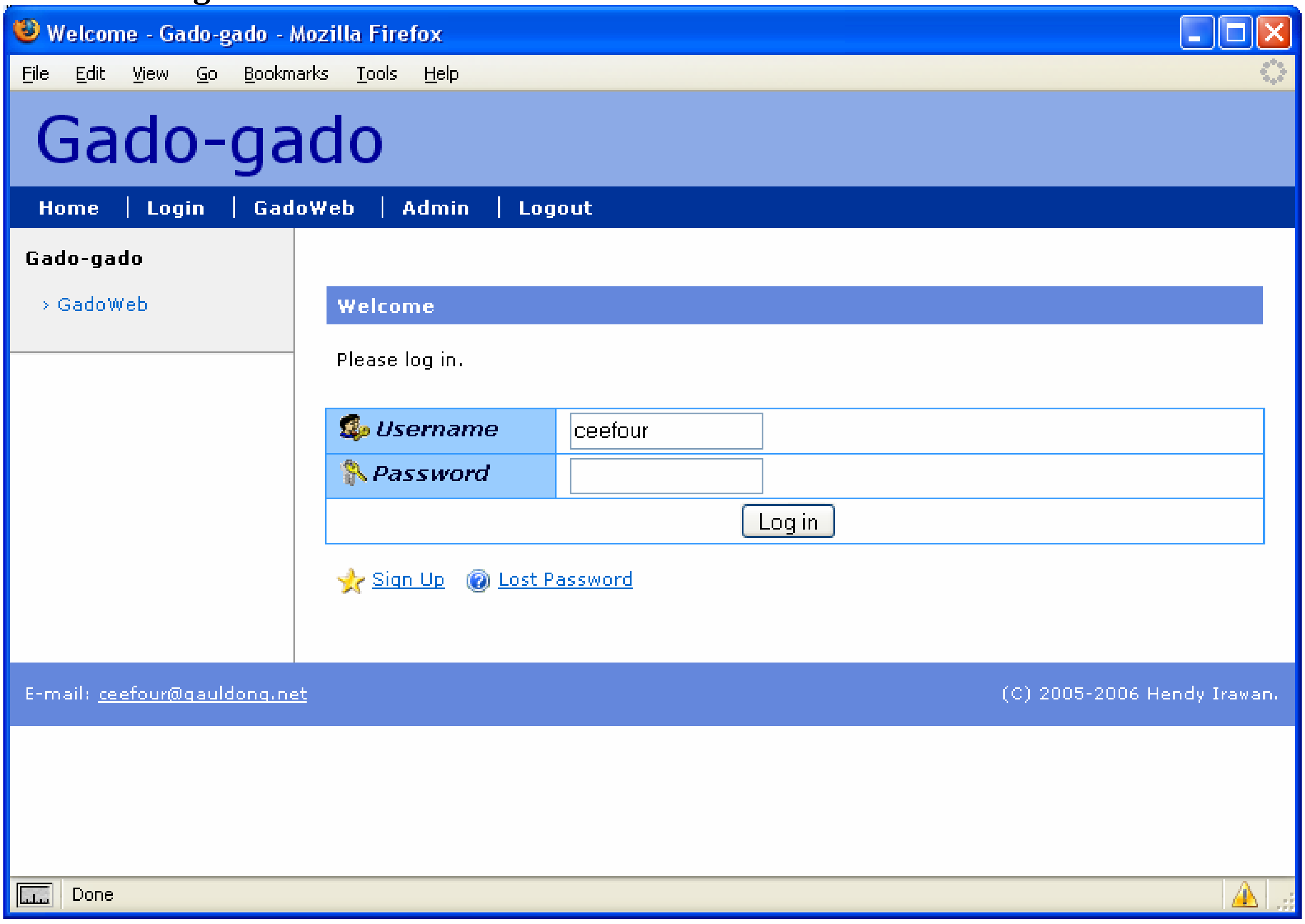

## Halaman Menu Utama

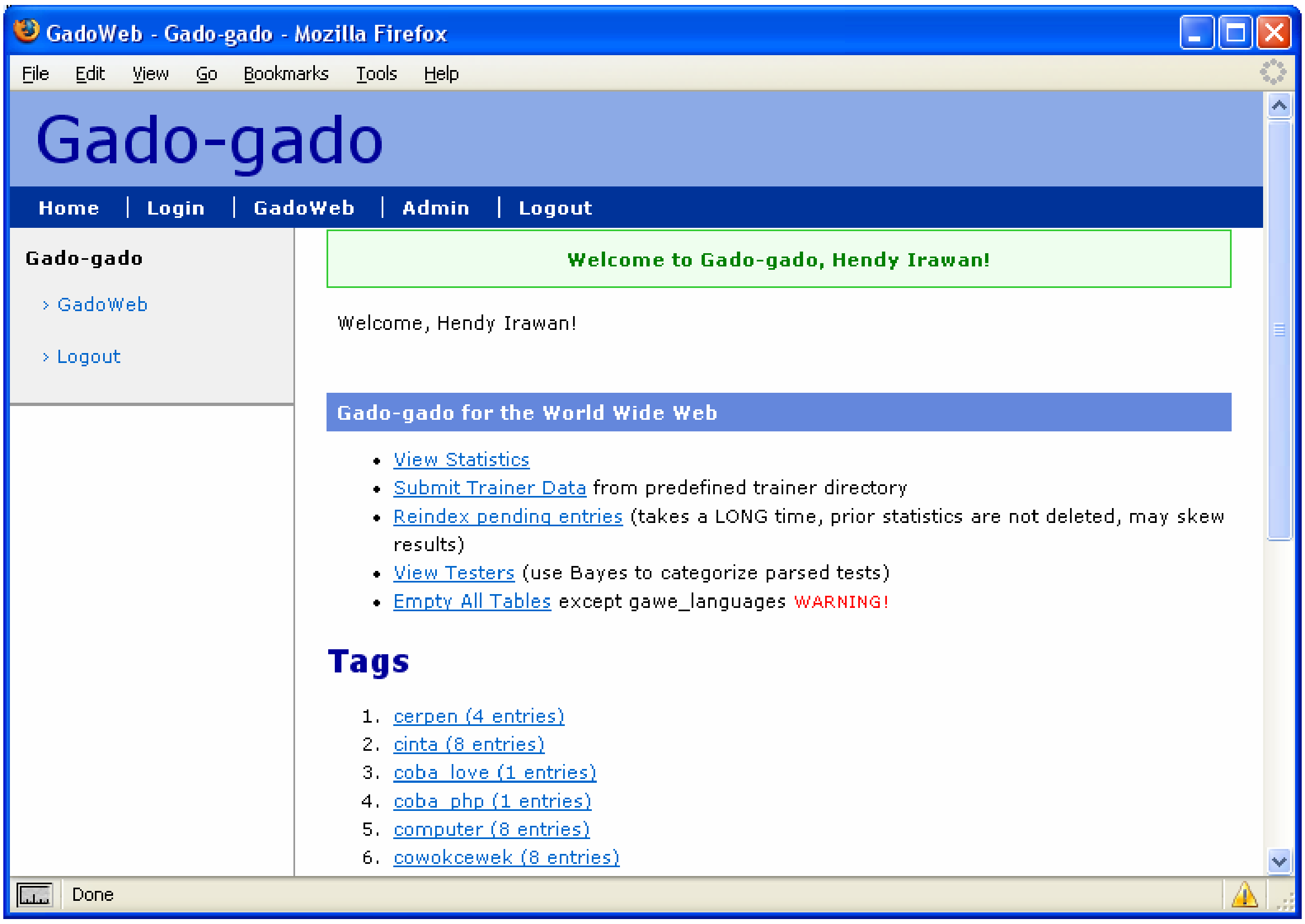

## Halaman Lihat Statistik

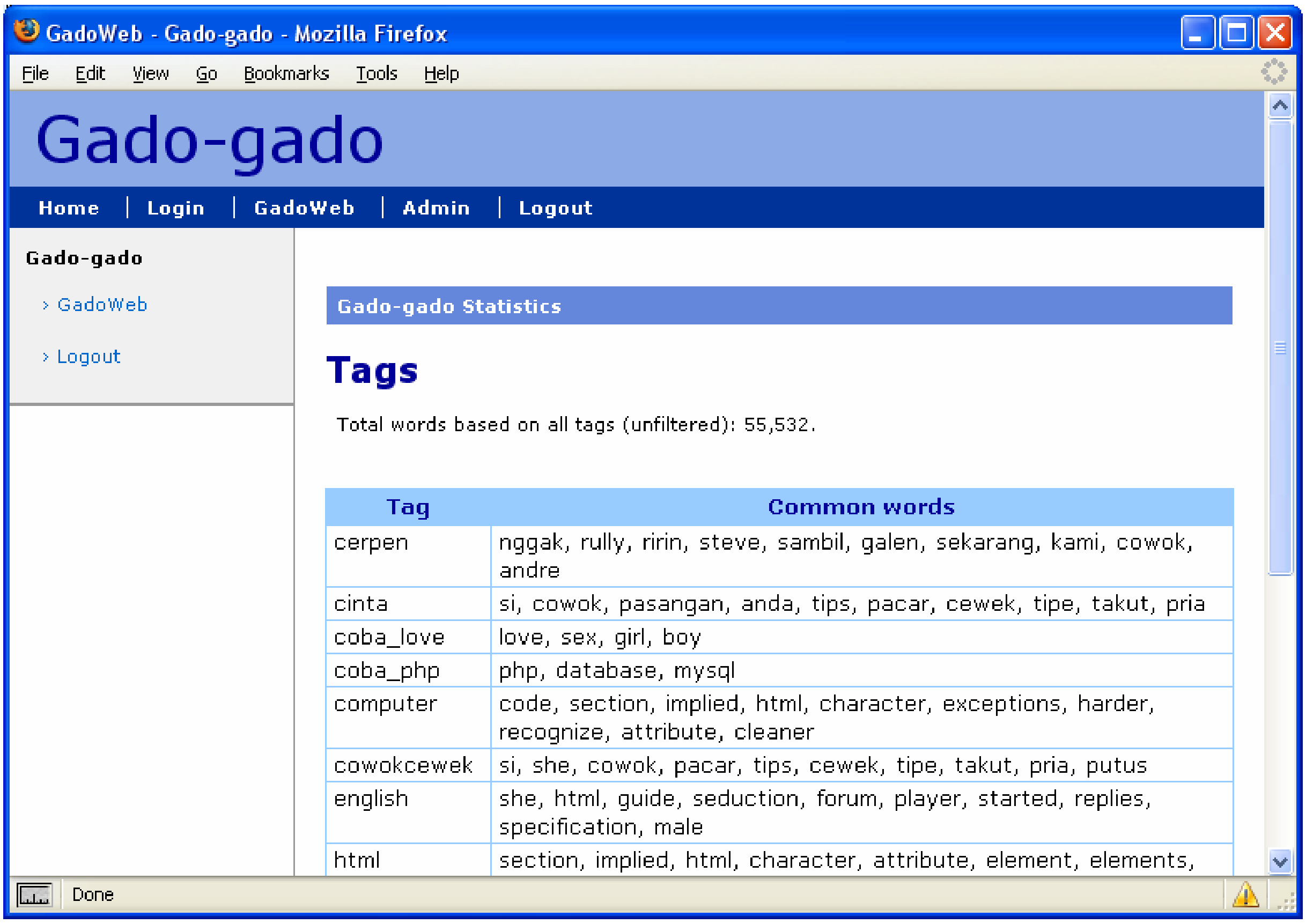

## Halaman Hasil Pengujian

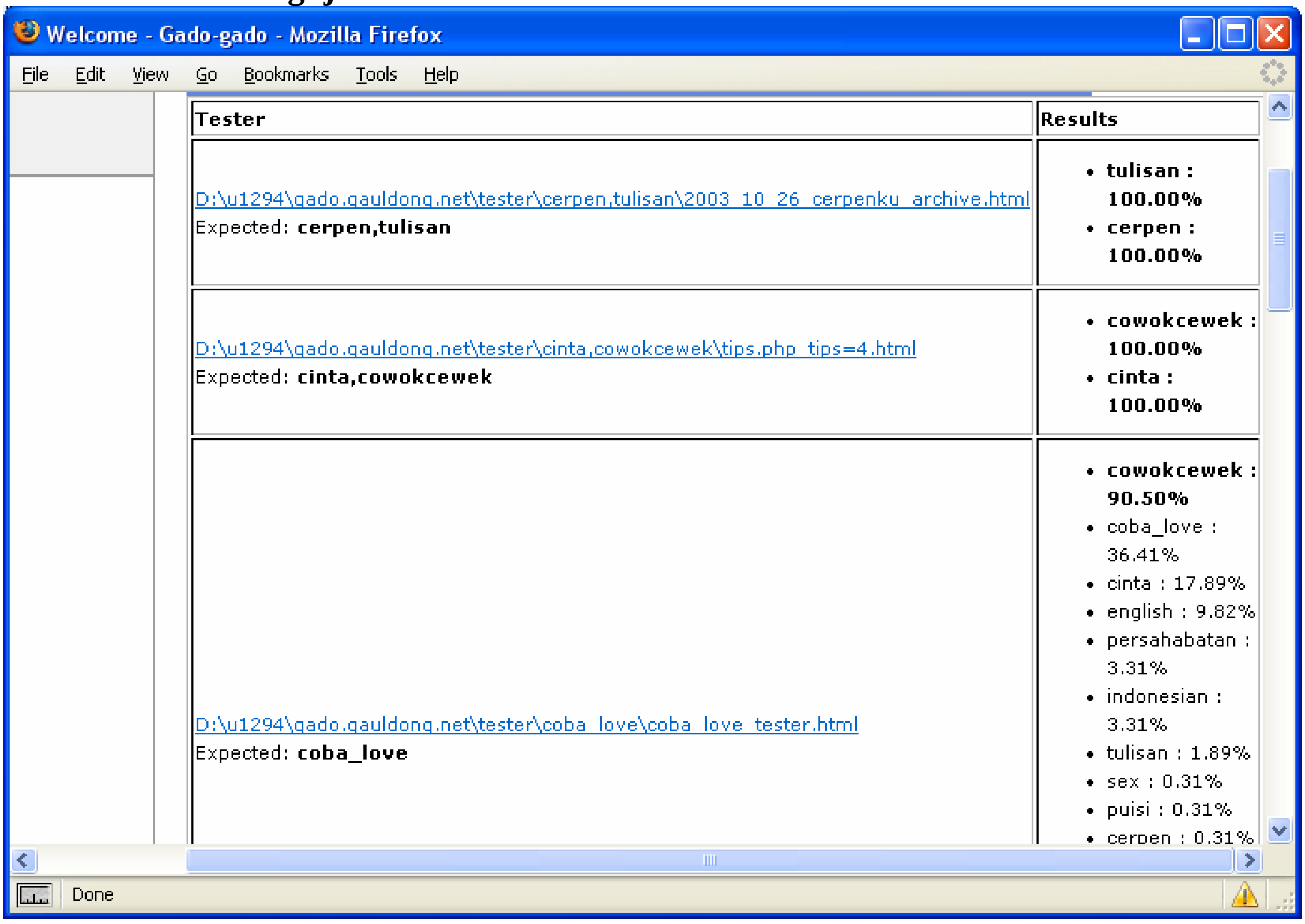

# Lampiran II
# Isi Tabel-tabel Database

**gawe_entries (42 record)**
Karena keterbatasan tempat, data yang ditampilkan hanya record awal saja.

| | |
|---|---|
| **entry_id** | 1 |
| **uri** | file:///D:\u1294\gado.gauldong.net\trainer\coba_love\coba_love_trainer.html |
| **title** | coba_love_trainer.html |
| **time_add** | 2005-12-11 14:09:41 |
| **time_update** | 2005-12-11 14:09:48 |
| **indexing_attempts** | 1 |
| **time_indexed** | 2005-12-11 14:09:48 |
| **indexing_time** | 0.771505 |
| **size** | 87 |
| **md5_hash** | 59a7883b3d18a2a3a714f282ae3ffe45 |
| **num_occurs** | 11 |
| **num_words** | 4 |
| **words** | a:4:{s:4:"love";i:5;s:3:"sex";i:2;s:4:"girl";i:2;s:3:"boy";i:2;} |
| **lang_code** | id |

**gawe_languages (2 record)**

| lang_code | language | num_entries | num_words |
|---|---|---|---|
| en | English | 360 | 3335 |
| id | Indonesian | 597 | 2591 |

**gawe_langwords (3313 record)**
Karena keterbatasan tempat, data yang ditampilkan hanya 10 record pertama.

| lang_code | word_id | num_entries | prob_entries | is_common |
|---|---|---|---|---|
| id | 1 | 6 | 0.0100502512562814 | 0 |
| id | 2 | 2 | 0.0033500837520938 | 0 |
| id | 3 | 4 | 0.0067001675041876 | 0 |
| id | 4 | 4 | 0.0067001675041876 | 0 |
| id | 8 | 3 | 0.0050251256281407 | 0 |
| id | 9 | 3 | 0.0050251256281407 | 0 |
| id | 10 | 3 | 0.0050251256281407 | 0 |
| id | 11 | 8 | 0.0134003350083752 | 0 |
| id | 12 | 4 | 0.0067001675041876 | 0 |
| id | 13 | 3 | 0.0050251256281407 | 0 |

**gawe_map (74 record)**
Karena keterbatasan tempat, data yang ditampilkan hanya 10 record pertama.

| entry_id | tag_id | relevance |
|---|---|---|
| 1 | 1 | 1 |
| 2 | 2 | 1 |
| 3 | 3 | 1 |
| 3 | 4 | 1 |
| 4 | 3 | 1 |
| 4 | 4 | 1 |
| 5 | 3 | 1 |
| 5 | 4 | 1 |
| 6 | 3 | 1 |
| 6 | 4 | 1 |

**gawe_subqueue (0 record)**
Tabel ini selalu dalam kondisi kosong, kecuali apabila ada dokumen baru yang ditambahkan.

**gawe_subs (42 record)**
Karena keterbatasan tempat, data yang ditampilkan hanya 10 record pertama.

| user_id | entry_id | comments | time_add |
|---|---|---|---|
| 1 | 1 | | 2005-12-11 14:09:41 |
| 1 | 2 | | 2005-12-11 14:09:41 |
| 1 | 3 | | 2005-12-11 14:26:17 |
| 1 | 4 | | 2005-12-11 14:26:17 |
| 1 | 5 | | 2005-12-11 14:26:17 |
| 1 | 6 | | 2005-12-11 14:26:17 |
| 1 | 7 | | 2005-12-11 14:28:17 |
| 1 | 8 | | 2005-12-11 14:28:17 |
| 1 | 9 | | 2005-12-11 14:28:17 |
| 1 | 10 | | 2005-12-11 14:28:17 |

**gawe_tags (14 record)**
Karena keterbatasan tempat, data yang ditampilkan hanya 10 record pertama.

| tag_id | tag | time_add | time_update | num_occurs | num_entries | word_num_entries |
|---|---|---|---|---|---|---|
| 1 | coba_love | 2005-12-11 14:09:48 | 2005-12-11 14:09:48 | 11 | 1 | 4 |
| 2 | coba_php | 2005-12-11 14:09:48 | 2005-12-11 14:09:48 | 9 | 1 | 3 |
| 3 | cerpen | 2005-12-11 14:26:32 | 2005-12-11 14:26:40 | 7088 | 4 | 1713 |
| 4 | tulisan | 2005-12-11 14:26:33 | 2005-12-11 14:39:27 | 7623 | 12 | 1880 |

| tag_id | tag | time_add | time_update | num_occurs | num_entries | word_num_entries |
|---|---|---|---|---|---|---|
| 5 | php | 2005-12-11 14:28:37 | 2005-12-11 14:28:45 | 8248 | 4 | 1456 |
| 6 | computer | 2005-12-11 14:28:40 | 2005-12-11 14:39:26 | 14362 | 8 | 2509 |
| 7 | cinta | 2005-12-11 14:31:05 | 2005-12-11 14:39:30 | 3090 | 8 | 901 |
| 8 | cowokcewek | 2005-12-11 14:31:06 | 2005-12-11 14:33:34 | 3576 | 8 | 1032 |
| 9 | english | 2005-12-11 14:33:31 | 2005-12-11 14:35:04 | 2923 | 8 | 843 |
| 10 | html | 2005-12-11 14:39:19 | 2005-12-11 14:39:25 | 6114 | 4 | 1053 |

**gawe_tagwords (7733 record)**
Karena keterbatasan tempat, data yang ditampilkan hanya 10 record pertama.

| word_id | tag_id | num_entries | prob_entries | num_occurs |
|---|---|---|---|---|
| 1 | 1 | 1 | 1 | 5 |
| 2 | 1 | 1 | 1 | 2 |
| 3 | 1 | 1 | 1 | 2 |
| 4 | 1 | 1 | 1 | 2 |
| 5 | 2 | 1 | 1 | 5 |
| 6 | 2 | 1 | 1 | 2 |
| 7 | 2 | 1 | 1 | 2 |
| 8 | 3 | 2 | 0.5 | 80 |
| 9 | 3 | 2 | 0.5 | 48 |
| 10 | 3 | 2 | 0.5 | 43 |

**gawe_words (3276 record)**
Karena keterbatasan tempat, data yang ditampilkan hanya 10 record pertama.

| word_id | word | num_entries | num_occurs |
|---|---|---|---|
| 1 | love | 5 | 13 |
| 2 | sex | 5 | 24 |
| 3 | girl | 7 | 66 |
| 4 | boy | 3 | 46 |
| 5 | php | 4 | 50 |
| 6 | database | 2 | 4 |
| 7 | mysql | 2 | 5 |
| 8 | galen | 2 | 80 |
| 9 | tante | 2 | 48 |
| 10 | mama | 2 | 43 |

**users (1 record)**

| | |
|---|---|
| **user_id** | 1 |
| **login** | ceefour |
| **passwd** | d41d8cd98f00b204e9800998ecf8427e |
| **name** | Hendy Irawan |
| **email** | ceefour@gauldong.net |
| **role_name** | Site Admin |
| **time_join** | 2005-02-01 00:00:00 |
| **time_login** | 2005-09-29 13:08:47 |
| **time_reset** | NULL |
| **time_online** | 2005-09-29 13:20:10 |
| **timezone** | Asia/Jakarta |
| **activated** | 1 |